\documentclass[fleqn,usenatbib]{mnras}
\usepackage{newtxtext,newtxmath} 
\usepackage[whole]{bxcjkjatype}
\usepackage[T1]{fontenc} 
\DeclareRobustCommand{\VAN}[3]{#2}
\let\VANthebibliography\thebibliography
\def\thebibliography{\DeclareRobustCommand{\VAN}[3]{##3}\VANthebibliography}
\usepackage{graphicx}	
\usepackage{amsmath}	
\usepackage{bm}  
\usepackage{color}   
\usepackage{mfirstuc}
\def\be{\begin{equation}}\def\ee{\end{equation}}
\def\vlsr{v_{\rm LSR}} \def\Msun{M_\odot} \def\vr{v_{\rm r}}
\def\deg{^\circ}  \def\vlsr{v_{\rm lsr}} \def\vrot{V_{\rm rot}}
\def\vlsr{V_{\rm lsr}} \def\Vrot{V_{\rm rot}}
\def\co{$^{12}$CO }  
\def\coth{$^{13}$CO } \def\coei{C$^{18}$O }
\def\Xco{X_{\rm CO}}   \def\Tb{T_{\rm B}} \def\Tp{T_{\rm p}}
\def\Htwo{H$_2$ }     
 \def\Msun{M_\odot}  
 \def\kms{km s$^{-1}$} \def\Ico{I_{\rm CO}}  
\def\mH{m_{\rm H}}  \def\Htwo{H$_2$ } \def\mum{$\mu$m }
\def\Ico{I_{\rm CO}}  \def\Tb{T_{\rm B}}  
 \def\mH{m_{\rm H}}     \def\ekms{{\rm km\ s^{-1}}} 
 
 \def\apj{ApJ} \def\aap{AA} \def\mnras{MNRAS} \def\pasj{PASJ} \def\aj{AJ}
 \def\xcounit{H$_2$ cm $^{-2}$ [K km s$^{-1}]^{-1}$}
 \def\log{{\rm log}} \def\Rgc{R_{\rm GC}} 
 \def\Htwo{H$_2$} \def\percc{cm$^{-3}$} 
 \def\htwcc{\Htwo \percc} 
 \def\Hii{ {H\small{II}} } 
 \def\HII{ {H\small{II}} } 
 \def\Hi{ {H\small{I}} }

\title[GMC G18.1-0.3 associated with HII regions and SNR]{Giant molecular cloud G18.1-0.3+51 associated with HII regions and supernova remnant in the 3-kpc expanding ring} 
 \author[Y. Sofue]{Yoshiaki Sofue \\ 
Institute of Astronomy, The University of Tokyo, Mitaka, Tokyo 186-0015, Japan}
\date{Accepted; Received YYY; in original form} 
\pubyear{2023}  
\begin{document} 
\maketitle 
\begin{abstract} 
Analyzing the high-resolution CO-line survey of the Galactic plane with the Nobeyama 45-m telescope (FUGIN), we show that the star-forming complex G18.15-0.30+51 (G18) at radial velocity of 51 \kms is {a tight triple association of a giant molecular cloud (GMC), \Hii regions, and a supernova remnant (SNR).}
The radial velocity of G18 allows three possible kinematic distances of $d=3.9\pm 0.2$ kpc for near solution or $12\pm 0.2$ kpc for far solution, if we assume circular Galactic rotation, or $d=6.1\pm 0.1$, if it is moving with the 3-kpc expanding ring at an expanding velocity of 50 \kms.
The\Hi-line absorption of radio continuum from the \HII region constrains the distance to $5.6 \lesssim d_{\rm SNR}\le 7.6$ kpc.
The $\Sigma-D$ (radio brightness-diameter) relation yields the distance to the SNR of $d_{\rm SNR}=10.1^{+11.5}_{-4.7}$ kpc, allowing for a minimum distance of 5.4 kpc.
From these we uniquely determined the distance of G18 to be $6.07\pm 0.13$ kpc in the 3-kpc expanding ring with the SNR being physically associated.  
The molecular mass of the GMC is estimated to be $M_{\rm mol}\sim 3\times 10^5 \Msun$.
The ratio of Virial to luminous molecular masses is greater than unity in the central region and decreases outward to $\lesssim 0.2$ at the cloud edge, indicating that the central region is dynamic, while the entire cloud is stable.
We discuss the origin of the G18 triple system and propose a sustainable GMC model with continuous star formation. 
\end{abstract}  
  
\begin{keywords} ISM: bubbles --- ISM: clouds --- ISM: kinematics and dynamics --- ISM: supernova remnant --- ISM: \Hii regions ---stars: formation \end{keywords} 

\section{Introduction} 
\label{sec1}

In the survey of molecular-gas shells associated with supernova remnants (SNR) in the Galactic disc using the FUGIN\footnote{Four-receiver-system Unbiased Galactic plane Imaging survey with the Nobeyama 45-m telescope} CO-line data, several molecular complexes associated with both star-forming region and SNR have been found \citep{2021ApJS..253...17S}.
The molecular gas complex G18.15-0.30+51 (hereafter G18) is one of such 'triple' systems, composing a tight association of a giant molecular cloud (GMC), \Hii regions, and an SNR \citep{2013MNRAS.433.1619P,2022ApJ...925...60D}.
Such a system will provide us with an opportunity to study the star formation process from the birth till death in a closed system. 

However, physical association of the SNR with the system is still controversial because of the uncertain distances:
Radial velocities in the molecular, \Hi, and \Hii recombination lines in emission are measured to be $\vlsr \sim 51-54$ \kms.
The kinematic distance to the \Hii regions and molecular cloud assuming the circular rotation has been obtained to be $\sim 4$ kpc for the near-distance solution because of the absence of \Hi absorption at the terminal velocity of the Galactic disc
 \citep{2006ApJS..165..338Q,2013MNRAS.433.1619P,2022ApJ...925...60D}.
On the other hand, the distance to the SNR has only been constrained to $\lesssim 8$ kpc from the absence of \Hi line absorption at the terminal velocity \citep{2014MNRAS.438.1813L}. 
One of the purpose of the present paper is put a more precise constraint on the distances to the objects, and examine if the apparently triple system is indeed physically associated with each other.

Besides the radio shells like the SNR and \Hii regions, bubble structures of molecular gas are found, and are interpreted as due to the outflow driven by star forming activity in the cloud center \citep{2013MNRAS.433.1619P}, where cloud collisions are thought to play a role to trigger the massive-star formation \citep{2022ApJ...925...60D}.
We also aim at exploring detailed morphological and kinematical properties with particular emphasis of the relation of the molecular bubbles to the complex GMC. 
We further try to construct an integrated view of the star formation and feedback mechanism in the G18 system.

In section \ref{sec2}, we investigate the kinematic and dynamic properties of the GMC in G18 complex based on the detailed analysis of FUGIN (Four-receiver system Unbiased Galactic plane Imaging survey with the Nobeyama 45-m telescope) CO line data \citep{2017PASJ...69...78U}.
In section \ref{sec3} we determine the distances of the vrious components in G18, taking account of the kinematic model of the Galaxy including the expanding 3-kpc ring.
In section \ref{sec4}, based on the new distance and analysis, we investigate the interstellar physics and dynamics of the complex.
In section \ref{sec5} we discuss the origin of the G18 system,  and propose a model in which G18 is explained by a gravitationally stable structure with star-forming activity sustained by efficient feedback of gas.
Section \ref{sec6} presents the conclusion of the work.  

\section{The Data and Maps}
\label{sec2}

\subsection{Data}

\begin{figure*} 
\begin{center}     
\includegraphics[height=5.2cm]{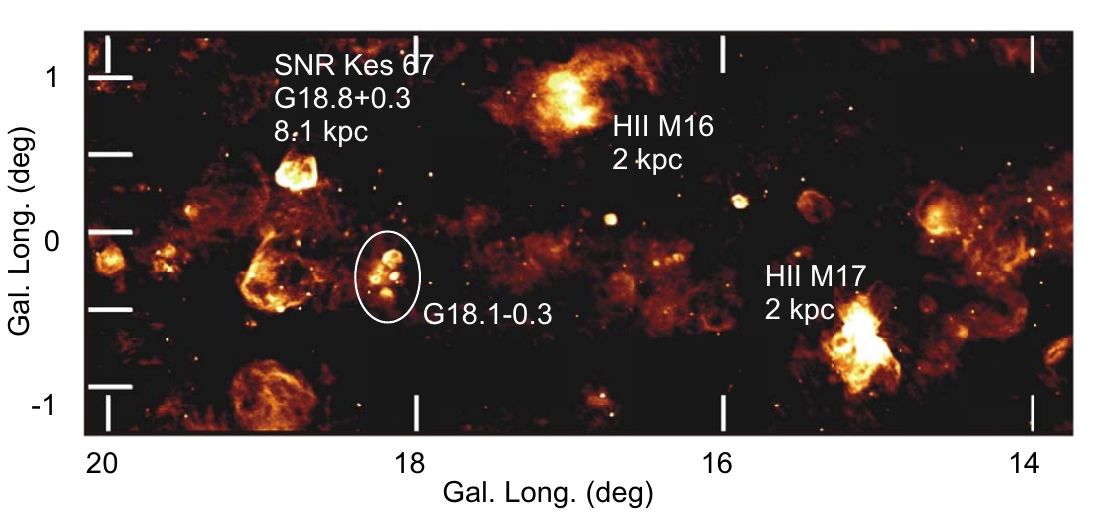} 
\includegraphics[height=5.2cm]{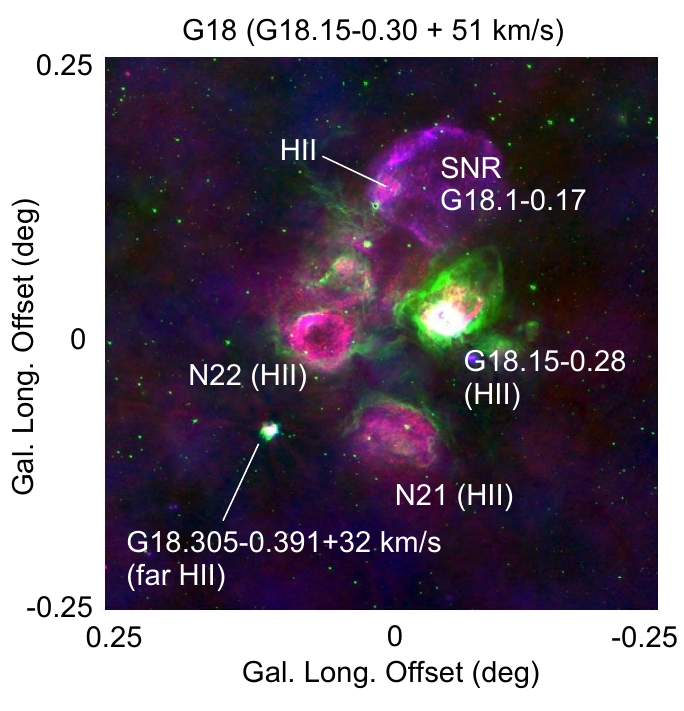}
\end{center}
\caption{
Galactic plane at 90 cm from the SNR survey by VLA (Brogan et al. 2006)
reproduced from NRAO News Letter at https://www.nrao.edu/archives/items/show/33519). The complex G18.1-0.3 is marked by the ellipse. The right panel shows $30' \times 30'$ area around G18 at 20 cm (red), 8\mum (green) and 90-cm (blue) extracted from MAGPIS (Haffner et al. 2006; Stil et al. 2006)
HII regions and an SNR are marked following (Paron et al. 2013)
{G18.305-0.391 is a background \Hii region at a distance 12.6 kpc (see Appendix).
}}
\label{fig1}
\end{figure*}

In order to obtain the general view of the region around G18, we show in figure \ref{fig1}
the radio continuum image a wide region around G18 at 90-cm taken with the VLA (Very Large Array) \citep{2006ApJ...639L..25B} and a composite map of far-infrared, 20-cm and 90-cm radio continuum survey maps extracted from the Multi-Array Galactic Plane Imaging Survey (MAGPIS) \citep{2006AJ....131.2525H}, and VLA Galactic Plane Survey (VGPS) \citep{2006AJ....132.1158S}.  

\begin{figure*} 
\begin{center}     
\includegraphics[height=9cm]{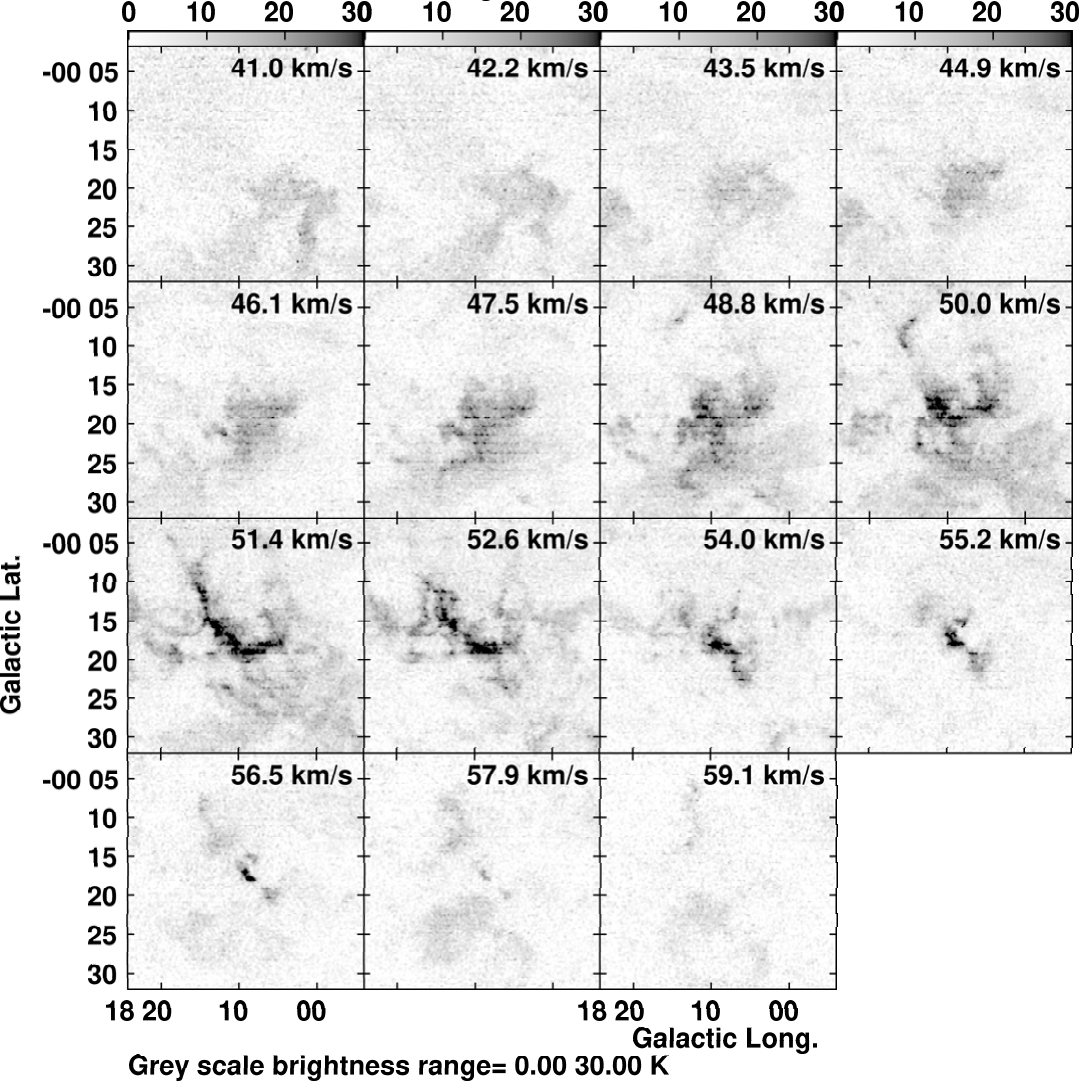}
\includegraphics[height=9cm]{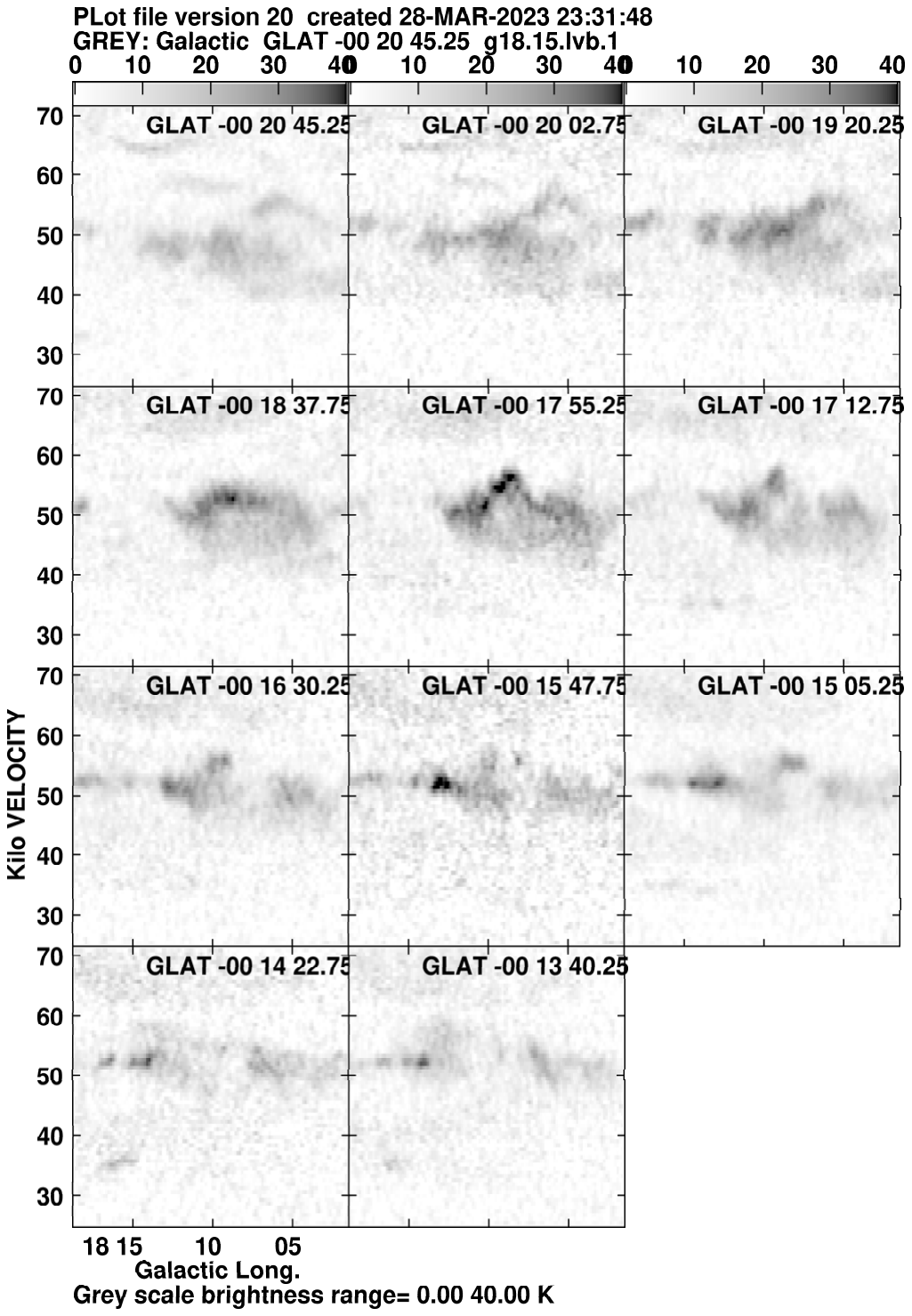}   
\end{center}
\caption{[Left] \co-line channel maps in grey scale from $\Tb=0$ to 30 K.
[Right] LVD across G18 at various latitudes. } 
\label{fig2}
\end{figure*}

For molecular cloud analysis, we made use of the FUGIN survey data \citep{2017PASJ...69...78U}.
The full beam width at half maximum of the telescope was 15\arcsec at the \co ($J=1-0$)-line frequency (115 GHz), and the velocity resolution was 1.3 \kms. 
The effective beam size of the final data cube is 20\arcsec, and the rms noise levels are $\sim 1$ K.  
The 3D FITS cubes have a pixel size of $(\Delta l, \Delta b, \Delta \vlsr) =$ (8''.5, 8''.5, 0.65 \kms). 
Fig. \ref{fig2} shows examples of the channel maps of \co-line brightness temperature ($\Tb)$ and longitude-velocity ($(l,\vlsr)$) diagrams at several latitudes and velocity channels.
  
\begin{figure} 
\begin{center}    
\includegraphics[height=8cm]{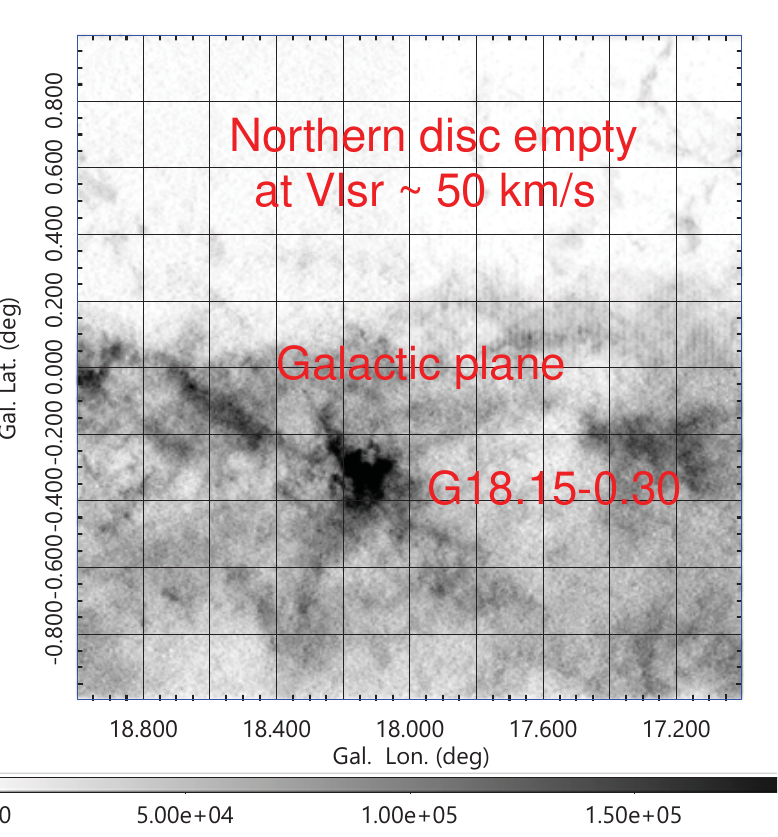} 
\end{center} 
\caption{ \co-line intensity map (moment 0) in $2\deg \times 2\deg$ around G18.0+0.0 integrated from $\vlsr=40$ to 60 \kms. 
Note the significant north-south asymmetry about the Galactic plane in the molecular gas distribution in this velocity range.  } 
\label{fig3}
\end{figure}    

\begin{figure*} 
\begin{center}     
\co  \hskip 5cm \coth \\ 
\includegraphics[height=6.7cm]{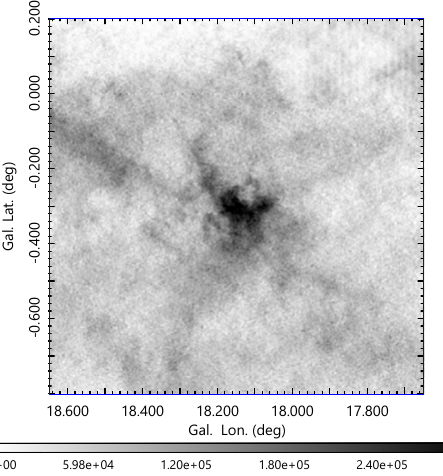}   
\includegraphics[height=6.7cm]{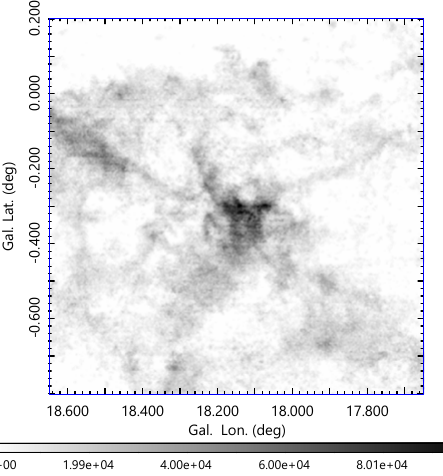}\\
\coei \hskip 4.5cm Composite \\
\includegraphics[height=6.7cm]{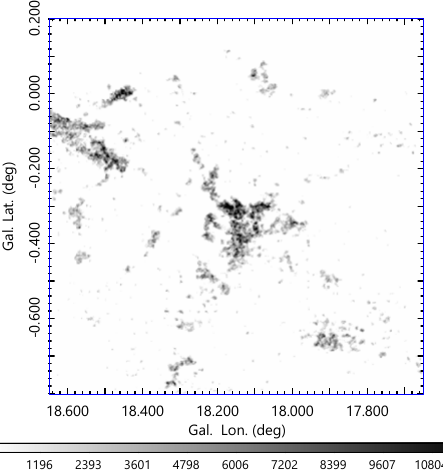}    
\includegraphics[height=6.7cm]{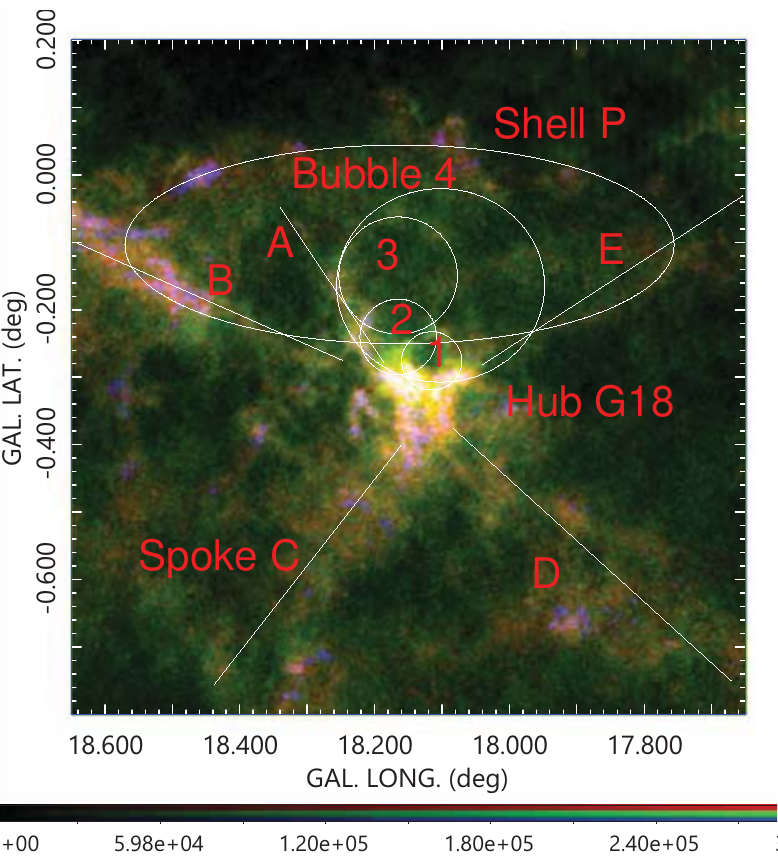}\\
\includegraphics[height=6.7cm]{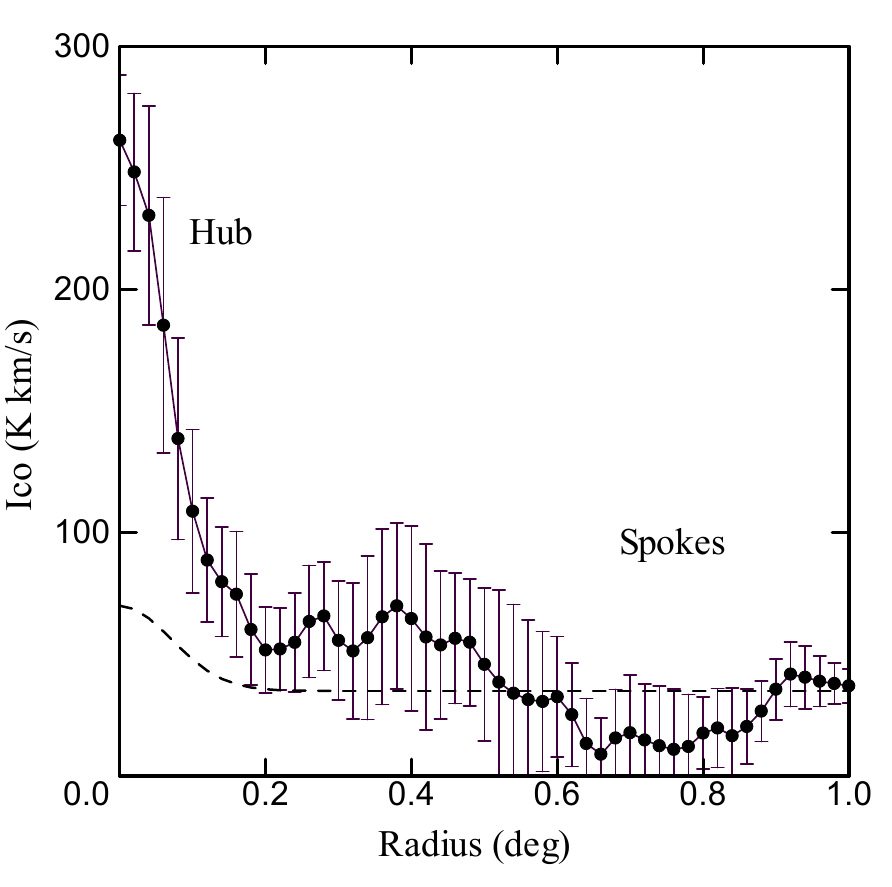}
\end{center} 
\caption{[Top left, right, middle left]
Moment 0 maps in the \co, \coth, \coei lines, respectively, for $1\deg\times 1\deg$ region.   
[Middle right]  Composite moment 0 map of the \co, \coth, \coei in RGB colors for $1\deg\times 1\deg$ around G18.15-0.30. 
Intensities are relative.
Spokes A to E, composing an 'X' structure centered on the hub at the cloud center, molecular bubbles (cavities) 1 to 4, and a big shell P, discussed in the text, are marked by white lines and circles.
Bubble 4 might not be an independent object, but might trace the outskirts of the inner three bubbles.
[Bottom] Averaged \co\ intensity as a function of the distance from the hub center along the spokes.
Background emission at $b\sim -1\deg$ is subtracted.
Dashed line shows a profile expected for two perpendicularly overlapped filaments with constant intensity.}
\label{fig4}   	 
\end{figure*}

\subsection{Intensity distributions}

Fig. \ref{fig1} illustrates the location of G18 in a 90-cm radio continuum survey map reproduced from the SNR survey with the VLA \citep{2006ApJ...639L..25B}.
G18 is marked by a circle, which is a compact association of several \Hii regions and an SNR in the northern top.
The right panel enlarges a $30'\times 30'$ area around $l=18\deg.15$, $b=-0\deg.5$ in 20 cm radio continuum by red, 90 cm by blue, and 8 \mum infrared emission by green color, as extracted from MAGPIS. 
 
Figure \ref{fig2} shows channel maps of the \co line brightness and longitude-velocity diagrams (LVDs) at various latitudes.
Figs. \ref{fig3} shows a wide area map of the \co\ intensity, showing an extremely lopsided distribution of the CO line emission in the observed region with respect to the Galactic plane, indicating that the molecular gas near this velocity, $\sim 50$ \kms, is almost empty in the northern side of the Galactic plane.

Figure \ref{fig4} shows intensity maps (moment 0 maps) around G18 integrated from $\vlsr= 40$ to 60 \kms in the \co, \coth, and \coei line emissions along with an overlay of these figures in RGB (red, green, blue) color. 
The bottom panel shows an intensity profile as a function of the distance from the hub center averaged among the major spokes.

\subsection{{Hub-filament/spoke-bubble structure of {the molecular gas}}}

High concentration of molecular gas toward G18 at $(l,b)=(18\deg.15, -0\deg.30)$ is seen as an isolated peak in the field, which we name it here the "hub" according to the terminology of the molecular cloud structure \citep{2009ApJ...700.1609M}.
Several molecular branches extend from the hub extending in the directions at $PA\sim 20\deg, 70\deg,\ 150\deg, \ 220\deg$ and $\sim 300\deg$, which we call "spokes" A, B, C, D, and E, respectively, as marked by the white lines in Figs. \ref{fig3} and \ref{fig4}.
{The system composes a structure often called the hub-filament system (HFS) of interstellar dust lanes \citep{2009ApJ...700.1609M,2020A&A...642A..87K}. }
We here introduce a term 'spoke' to represent the constructive role of the giant and straight structures in conjunction to the 'hub', while it has the same meaning as 'giant filament' in the filament paradigm
 \citep{2022arXiv220309562H,2022arXiv220503935P}.

The spokes compose a large X-shaped structure extending over $\pm 0\deg.6$.
Spoke A is densest but shortest, composing multiple cavity and {molecular bubble} (shell) structures.
Spoke B extends almost straightly for more than $\sim 0\deg.6$ toward NE, making a counter part to the also straight spoke D to the SW.
Spokes B and E approximately trace the southern rim of the big molecular shell noticed in the earlier study marked as Shell P \citep{2013MNRAS.433.1619P}.
 
\begin{figure} \begin{center}      
\includegraphics[height=6.7cm]{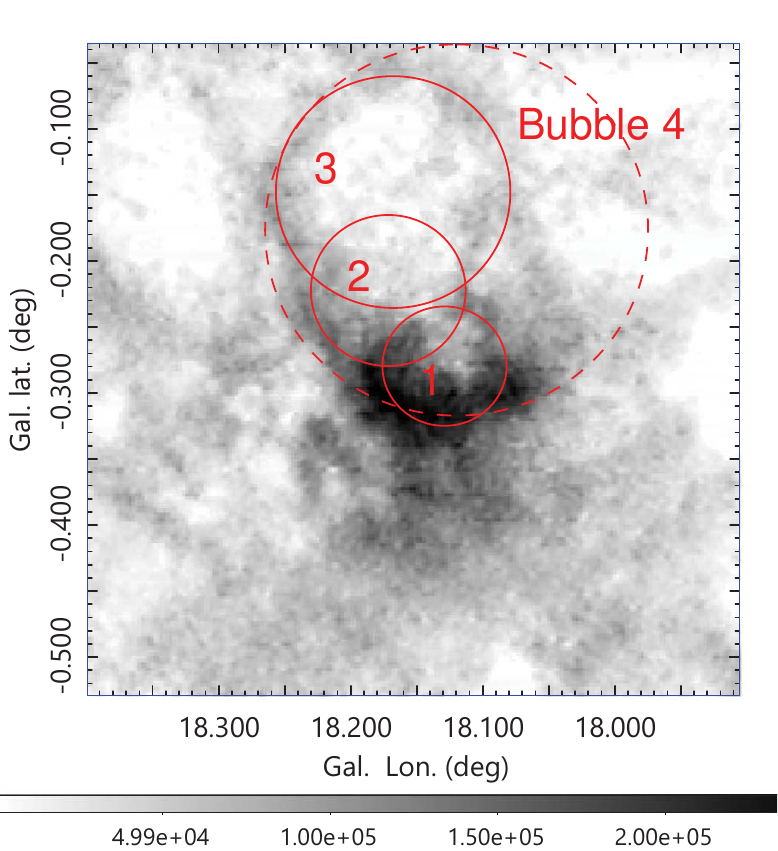}    
\includegraphics[height=6.7cm]{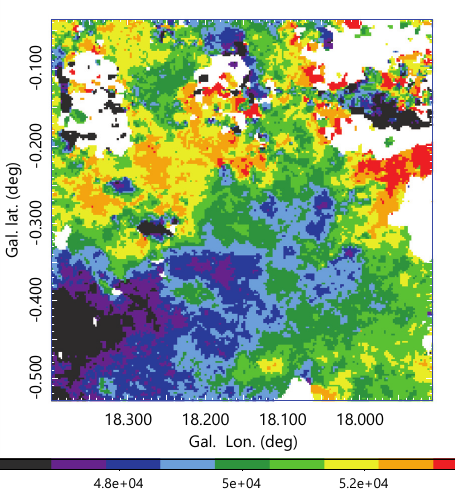}  
\includegraphics[height=6.7cm]{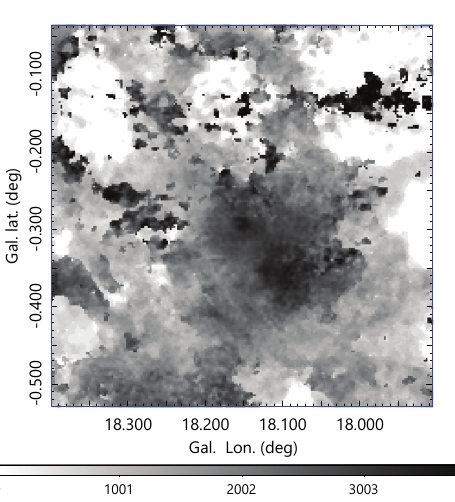}    
\end{center}
\vskip -2mm
\caption{[Top] \co-line moment 0 map integrated from $\vlsr=45$ to 55 \kms (grey bar scale is K m s$^{-1}$) with the molecular bubbles marked by red circles.
The cut-off level for integration is set to $\Tb \le 5$ K.
[Middle] Moment 1 map (velocity field)  in m s$^{-1}$.
[Bottom] Moment 2 map (velocity dispersion) in m s$^{-1}$. 
White regions are where no emission brighter than 5 K (cut off level) is detected.}
\label{fig5}   
\end{figure}

Spoke A toward the NW is connected to several open cavities, as marked by the white circles, which we name the {molecular bubbles} 1, 2, 3 and 4.
The molecular bubbles are more clearly seen in the moment 0 map shown in figure \ref{fig5}(a) for the core $30'\times 30'$ area, where the cut-off brightness was set to $\Tb\le 5$ K.

\begin{figure}
\begin{center} 
\includegraphics[height=6.7cm]{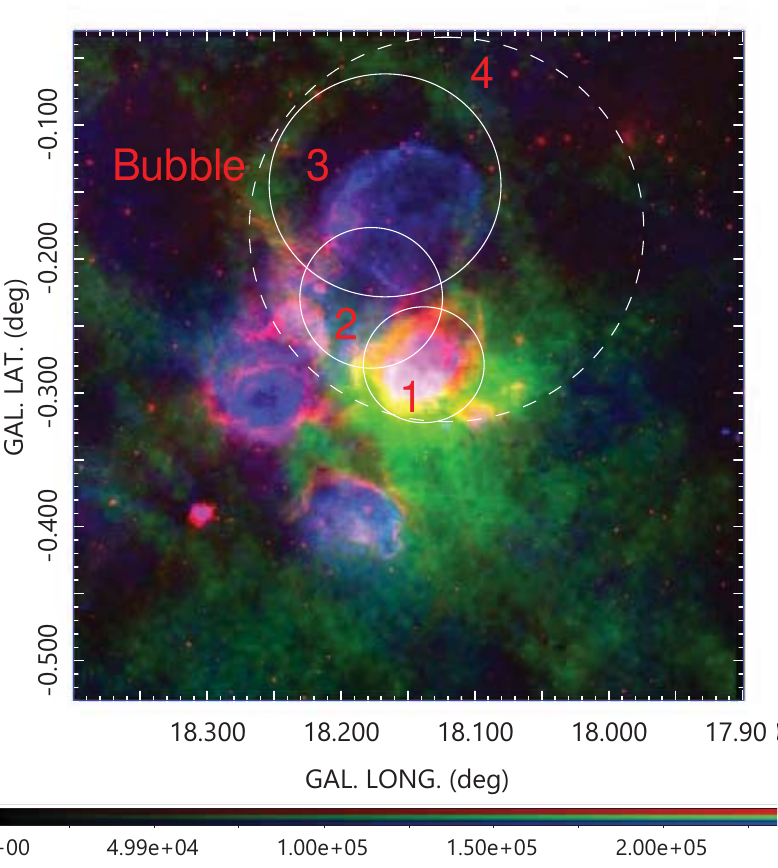}  
\includegraphics[height=6.7cm]{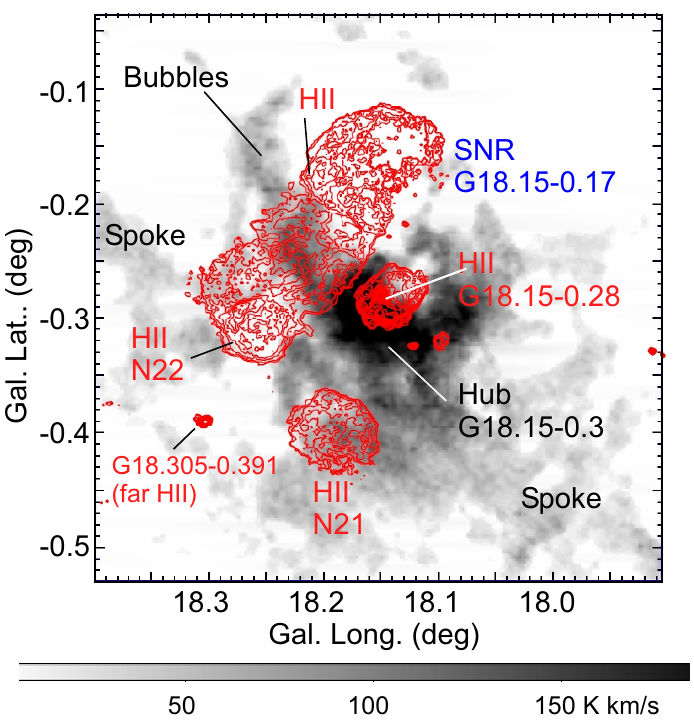}  
\includegraphics[height=6.7cm]{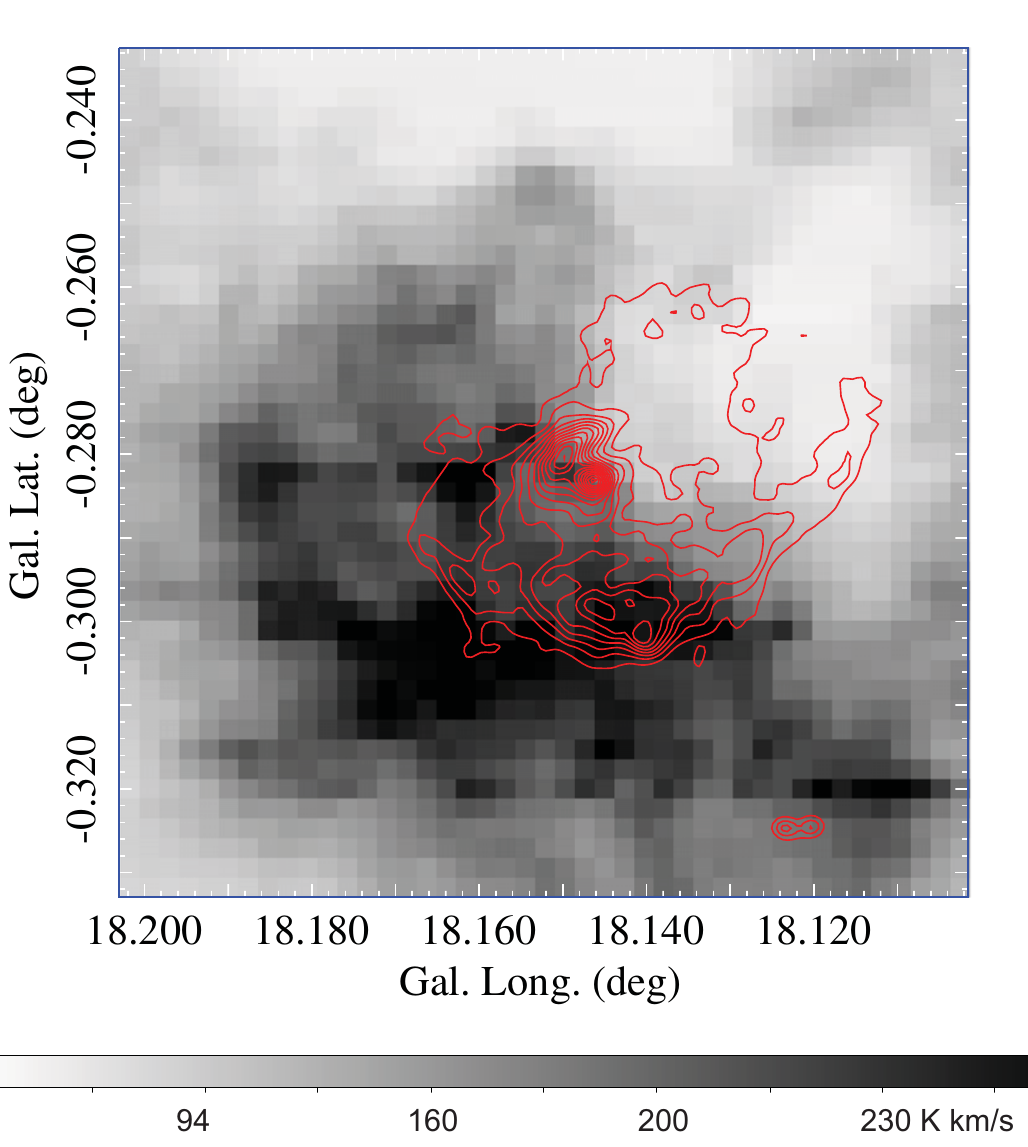}   
\end{center} 
\vskip -2mm
\caption{ 
[Top] Overlay of 8$\mu$m intensity in red, \co\ in green, and 20-cm continuum in blue. 
[Middle] \co intensity overlaid by 20-cm contours every 1 mJy beam$^{-1}$ with the lowest at 2.
[Bottom] Same, but {molecular bubble} 1 is enlarged superposed with 20-cm contours every 5 mJy beam$^{-1}$. The \Hii hot spot G18.147-0.280 is located at the bubble centre, while the \Hii shell traces the inner wall of the {molecular bubble}.} 
\label{fig6}  
\end{figure}

Molecular bubble 1 (hereafter, "bubble") is almost perfect, and the bubble's center coincides in position with the strong radio continuum source (spot) of the \Hii region G18.146-0.280 as shown in figure \ref{fig6} (b).
The inner edge of the bubble exactly traces the southern rim of the \Hii shell surrounding the radio spot.
Their coherent and concentric orientation strongly support the picture that they are formed by caving compression of the molecular hub by a high pressure \Hii sphere produced around the hot spot, possibly nesting O stars \citep{2013MNRAS.433.1619P}.

{Bubbles} 2 and 3 coincide with fainter \Hii regions, apparently overlapping with the western rim of SNR G18.1-0.1 on the same sight line.
{Bubble} 4 encircles the entire \Hii region and SNR area, which might not be an independent object from the others.

The multiple {bubbles} with their intensity getting fainter according to their increasing sizes and distances from the GMC center (hub) suggests a sporadic injection of kinetic energy in the core region of the GMC along the scenario of sequential star formation \citep{1977ApJ...214..725E,2013MNRAS.433.1619P}.
  
\subsection{Kinematics}

The moment 1 map (velocity field) in figure \ref{fig5} shows a velocity gradient from the south (blue shifted) to north (red shift).
This indicates either large-scale rotation on the SN plane, or an outflow from the center toward the north and away from the Sun.

The moment 2 map shows increasing velocity dispersion toward the hub center, attaining the highest value near the center at two positions.
The NW peak of the velocity dispersion coincides with the CO intensity peak and the brightest spot of the \Hii region \citep{2013MNRAS.433.1619P}.

Note that the moment 2 map here is biased to indicate smaller velocity dispersion due to the cut-off brightness at $\Tb\le 5$ K, which suppressed extended low brightness emission with broader velocity width.

\section{The Distance}
\label{sec3}
 
The kinematical distance for circular rotation curve has been derived to be $d\simeq 4$ for the near-side solution and $\sim 14$ kpc for far-side corresponding to the Galacto-centric distance of $R_{\rm GC}\simeq 6$ kpc  \citep{1980A&AS...40..379D}. 
The near/far ambiguity has been resolved using the absence of \Hi absorption of the radio continuum emission of the \Hii region at the terminal velocity of 130 \kms, yielding the near distance of $d=4.1$ \citep{2006ApJS..165..338Q} to $4.2$ kpc \citep{2009ApJS..181..255A}.
In this section, we newly determine the distances to the GMC, \Hii region, and SNR applying the kinematic method to the CO line, radio recombination line (RRL),\Hi absorption lines, and $\Sigma-D$ (surface brightness-diameter) relation to the SNR G18.15-0.17.
We particularly consider the non-circular rotation of the Galaxy in the 3-kpc expanding ring (3-kpc ER). 

\subsection{Distance of G18 in circular rotation}     

Fig. \ref{fig8}  shows LVD along the Galactic plane from the Columbia \co line survey \citep{2001ApJ...547..792D} and an LVD at $b=-0\deg.3$ made from the FUGIN \co survey.
Two major Galactic structures are indicated by dashed lines: the 4-kpc molecular ring (4-kpc MR) and 3-kpc ER drawing an inclined ellipse with the expanding velocity of 50 \kms at $l=0\deg$.
The molecular complex G18 is located at $(l,v)\sim (18\deg, +51)$ \kms) near the intersection of the 4-kpc MR and the 3-kpc ER.
The FUGIN LVD demonstrates how G18 is compact and particular in the plotted area.
There are three possible solutions about the kinematic location of G18 on the LVDs.
One and second are the near- and far-side arms on the 4-kpc MR in ordinal circular rotation, and the other is the approaching side of the 3-kpc ER.
 
Using the CO-line profiles shown in figure \ref{fig10} we measured the radial velocity of the molecular cloud to be $\vlsr=51\pm 2$ \kms.
This velocity agrees with that measured for the hydrogen recombination lines H$110\alpha$ of 52 to 53 \kms \citep{1980A&AS...40..379D,2006ApJS..165..338Q}.
 
\begin{figure} 
\begin{center}     
\includegraphics[width=8cm]{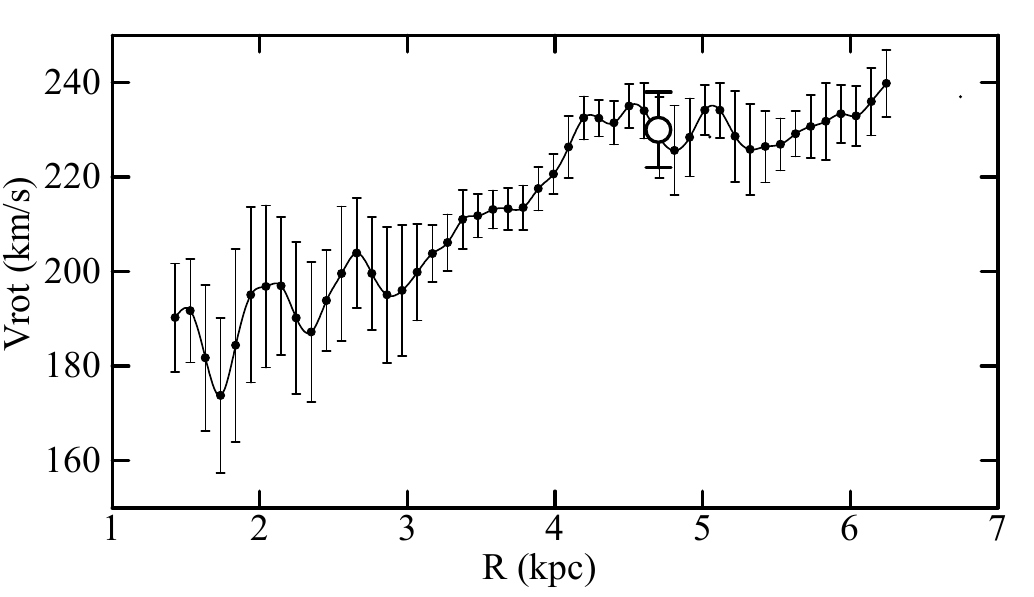} 
\end{center}
\caption{The most accurate rotation curve of the inner Milky Way in the first quadrant constructed from the FUGIN CO-line terminal velocities (Sofue 2021). 
Big circle indicates the Galacto-centric distance and rotation velocity of G18 obtained by iteration. 
The bar to the circle represents input error of the rotation velocity, considering the intrinsic fluctuation and errors of RC.}
\label{fig7} 
	\end{figure}

We here use the most accurate rotation curve (CO)  of the first quadrant of the inner Milky Way as shown in figure \ref{fig7} \citep{2021PASJ...73L..19S}.
The radial velocity $\vr$ and distance $d$ to the object in circular rotaiton are related by
\be
\vr=\left(\frac{R_0}{R}V(R)-V_0 \right) \sin\ l, 
\ee
and
\be
d=R_0 \cos\ l \pm \sqrt{R^2-R_0^2 \sin^2 l} \label{eqdis},
\ee
where $V(R)$ is the rotation velocity at Galacto-centric distance $R$, $V_0=238$ \kms, and $R_0=8.2$ kpc is the Galactic constants.
By iteration we obtain
$R=4.7\pm 0.2$ kpc,
and the near side solution of $d$ is
\be d_{\rm near}=3.9\pm 0.2 {\rm kpc},
\ee
and a far side solution 
\be 
d_{\rm far}=11.7\pm 0.2 {\rm kpc}.
\ee 
Because of the absence of the \Hi line absorption against continuum of the \Hii region, as described in the previous subsection, we may choose here the near solution of $d=3.9\pm 0.2$ kpc for the circular rotation. 

\subsection{Distance of G18 in the 3-kpc expanding ring}

As shown by the LVDs, G18 is located at the intersection of LVD ridges.
So, another possibility is that G18 is located on the 3-kpc ER.
If we assume that the source is moving with the expanding ring at $V_{\rm expa}$ and rotating at $\vrot$ obeying the RC, we have 
\be
\vr=\left(\frac{R_0}{R}V(R)-V_0 \right) \sin\ l \pm V_{\rm expa}\cos (l+\theta _{\pm}),
\ee
where $\theta_\pm$ is the GC angle satisfying
\be
d_{\pm} \cos \ l \pm R\cos \ \theta_\pm =R_0.
\ee
For the approaching motion of G18, we can attribute it in the near side of the 3-kpc ER and uniquely solve the equation to yield
\be
d=R_0 \cos \ l - \sqrt{\Rgc^2 - R_0^2 \sin ^2 l}.
\ee
Inserting $\Rgc=R_0 \sin \ l_{\rm tan}=3.07\pm 0.14$ pc with $l_{\rm tan}=22\deg \pm 1\deg$ as read by an ellipse fit to the 3-kpc ER's LV ridge in figure \ref{fig8} , we obtain 
\be
d=6.07 \pm 0.13\ {\rm kpc}.
\ee

\subsection{Intersection of 4-kpc molecular ring and 3-kpc expanding ring}

\begin{figure} \begin{center}  
(a)\includegraphics[width=7cm]{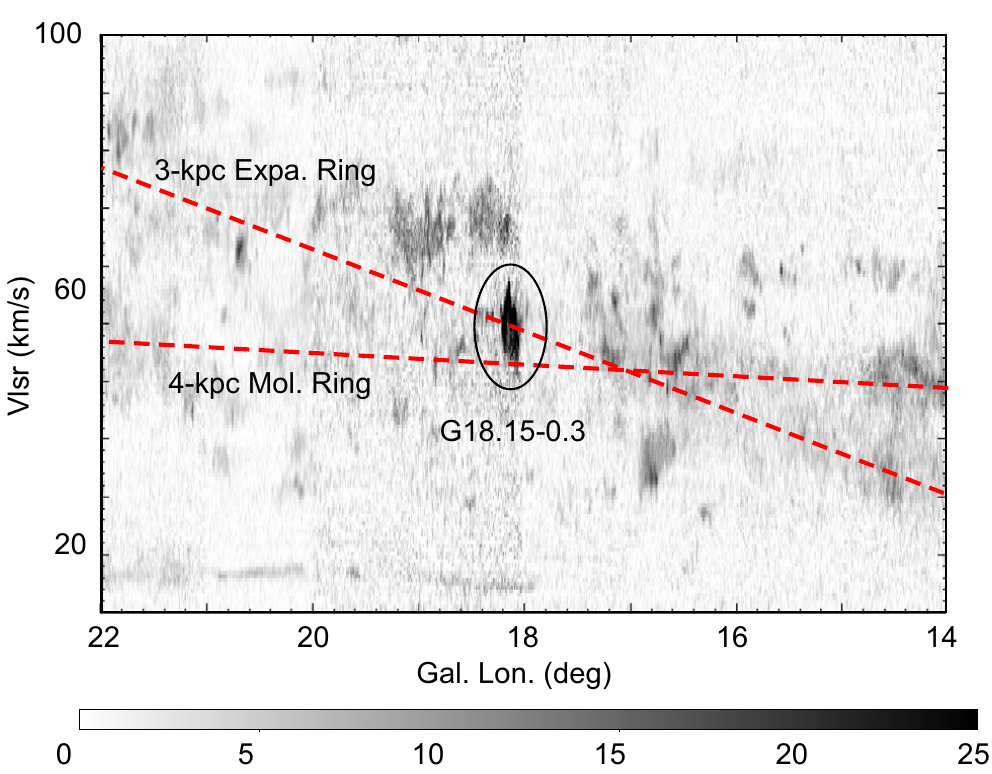}   \\
(b)\includegraphics[width=6.5cm]{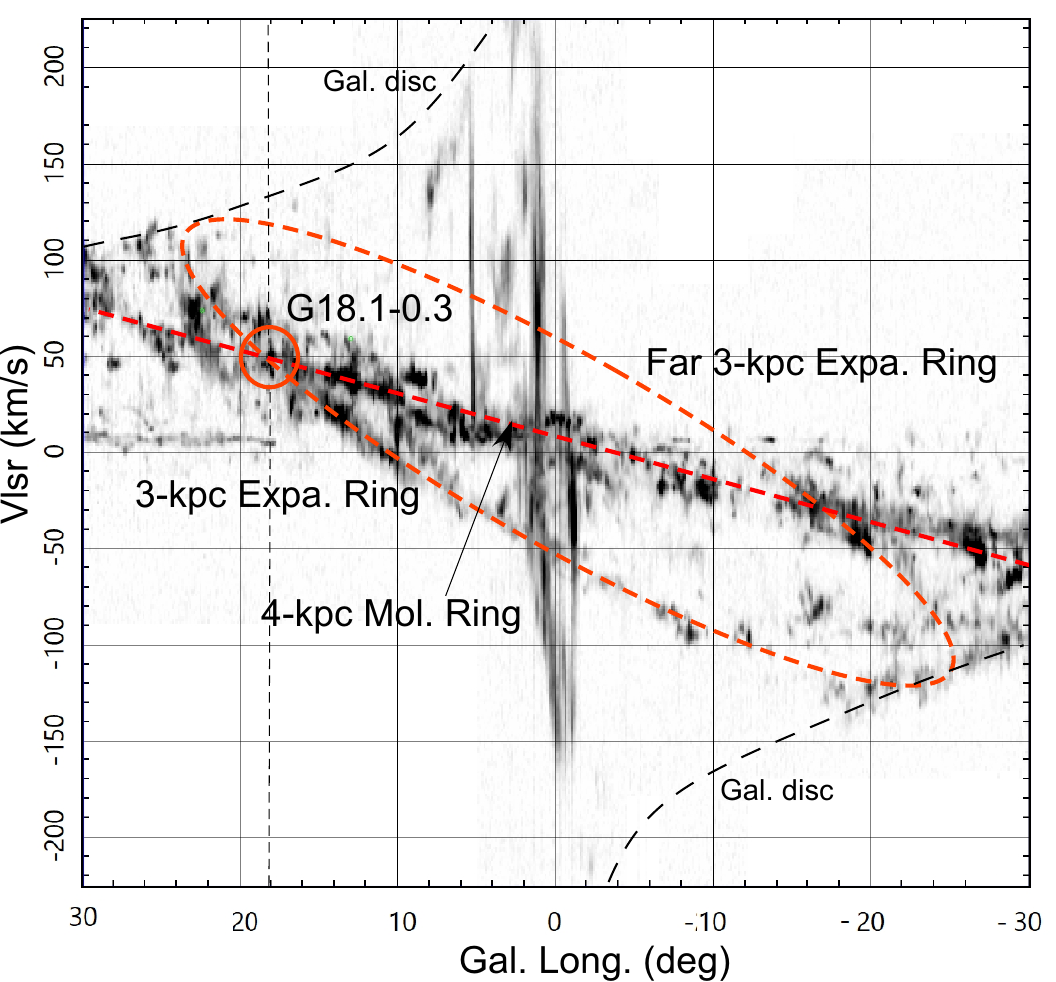}    \\
(c)\includegraphics[width=6cm]{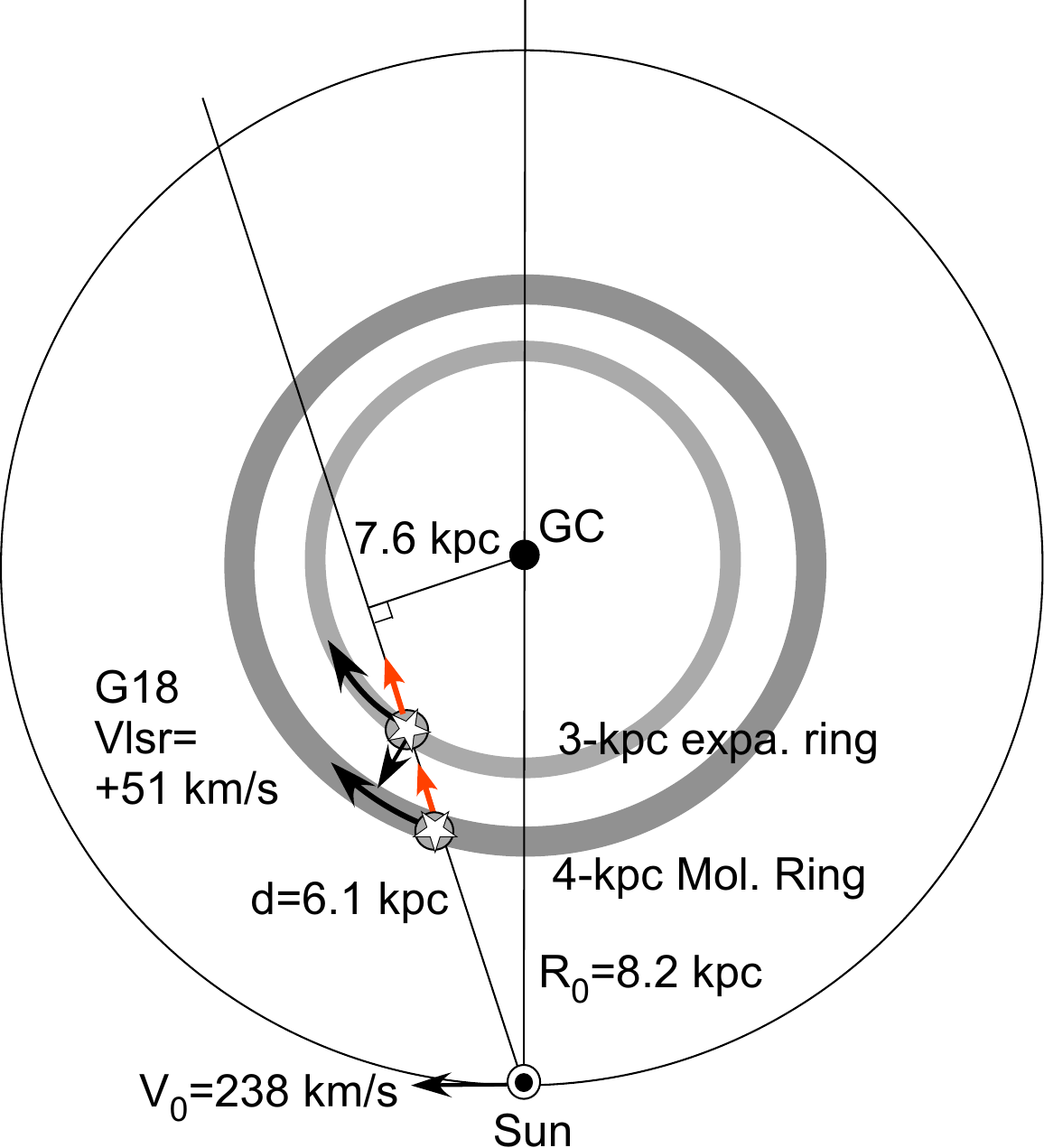}  
\end{center}
\caption{(a) FUGIN LVD at $b=-0\deg.3$ from $l=14\deg$ to $22\deg$.
Note that G18 is located near the intersection of the 3-kpc ER and 4-kpc MR.    (b) Colombia LVD at $b=-0\deg.375$ in the \co-line survey (Dame et al. 2001).
(c) Illustration of the mutual positions of the structures in the Galactic plane.} 
\label{fig8}  	\end{figure}

Fig. \ref{fig8}  shows a longitude-velocity diagram of the \co-line emission at $b=-0\deg.3$ in the central disc at $30\ge l \ge -30\deg$.
The two arm structures are evident:
One is the 4-kpc MR composed of dense molecular complexes, running straightly from $(l,v)\sim (-30\deg,-50 \ekms)$ to $(+30\deg,70\ekms)$, as marked by the red dashed line. 
The straight LV behavior indicates a circular rotating ring.
The other is the 3-kpc ER, composing a tilted ellipse on the LV plane, as traced by the red dashed line, which represents an expanding ring of radius 3 kpc and expansion velocity 50 \kms. 
It is stressed that G18 is located near the intersection of the two rings on the LV plane.
 
\subsection{{Distance of SNR and physical association with G18}}
 
The $\Sigma-D$  (surface brightness-diameter) relation at 1 GHz has been obtained for Galactic SNRs with known distances \citep{1998ApJ...504..761C,2013ApJS..204....4P}. 
We here adopt the most recent relation by \citep{2013ApJS..204....4P}
\begin{equation} 
\Sigma_{\rm 1GHz}=3.89^{+12.81}_{-2.98}\times 10^{-15} (D/1{\rm pc})^{-3.9\pm0.4}  
{\rm w~ m^{-2} Hz^{-1} sr^{-1}},
\end{equation} 
where $\Sigma_{\rm 1 GHz}$ and $D$ are the surface radio brightness at 1 GHz and linear diameter of the SNR, respectively. 
{In order to make the contamination by the thermal emission from the \Hii regions overlapping the SNR eastern shell as small as possible, we use here the lower frequency (90-cm continuum) map from VGPS survey  \citep{2006AJ....132.1158S}.
We measure the mean brightness on the map and converted it to that at 1 GHz assuming a spectral index of $\alpha=-0.6$. 
Using the thus obtained $\Sigma_{\rm 1\ GHz}$, we  determined the linear diameter by applying the $\Sigma-D$ relation to obtain 
$D=20.1^{+22.9}_{-9.3}$ pc and the distance 
\begin{equation}
 d_{\rm SNR}\sim 10.1^{+11.5}_{-4.7} {\rm kpc}. 
 \end{equation}  
 This distance locates the SNR farther than G18 molecular complex.
However, the large error allows for a possible minimum distance of $d\sim 10.1-4.7=5.4$ kpc. }

{On the other hand the\Hi-line absorption measurement toward the SNR has shown that the continuum radio emission from the SNR G18.15-0.17 exhibits absorption at $\vlsr \simeq 55$ and 100 \kms, but no absorption is found at the terminal velocity 130 \kms \citep{2014MNRAS.438.1813L}.
The absorption at 55 \kms is attributed to the 3-kpc ER at 6.1 kpc.
The 100 \kms absorption is due to the \Hi disc arm in circular rotation in front of the 3-kpc ER, and the kinematic distance is determined to be $5.6\pm 0.4$ kpc as the near solution, in good agreement with the current determination using\Hi-line absorption \citep{2014MNRAS.438.1813L}.
Another constraint is given by the absence of absorption at the terminal velocity, indicating that the SNR is located nearer than the tangent point at $\le 7.6$ kpc.}

{From these estimations, we can give a firm constraint on the distance of the SNR as $4.0 \le d_{\rm SNR} \le 7.6$ kpc.
This means that it is reasonable to locate the SNR in the 3-kpc ER at a distance of 6.07 kpc. 
Besides the distance, the apparently compact location on the sky of the SNR in touch with the \Hii regions and encircled by the molecular shells (Figs. \ref{fig6}) as well as the isolation of the entire G18 system on the sky (Fig. \ref{fig1}) supports the idea that the SNR is physically associated with G18. 
In more detail, the sharp \Hi absorption at 55 \kms \citep{2014MNRAS.438.1813L} indicates that the SNR is located in the far side of G18 as illustrated in figure \ref{fig8}. 
}
 
\subsection{The reliable distance}

\begin{figure}
\begin{center}   
\includegraphics[width=8cm]{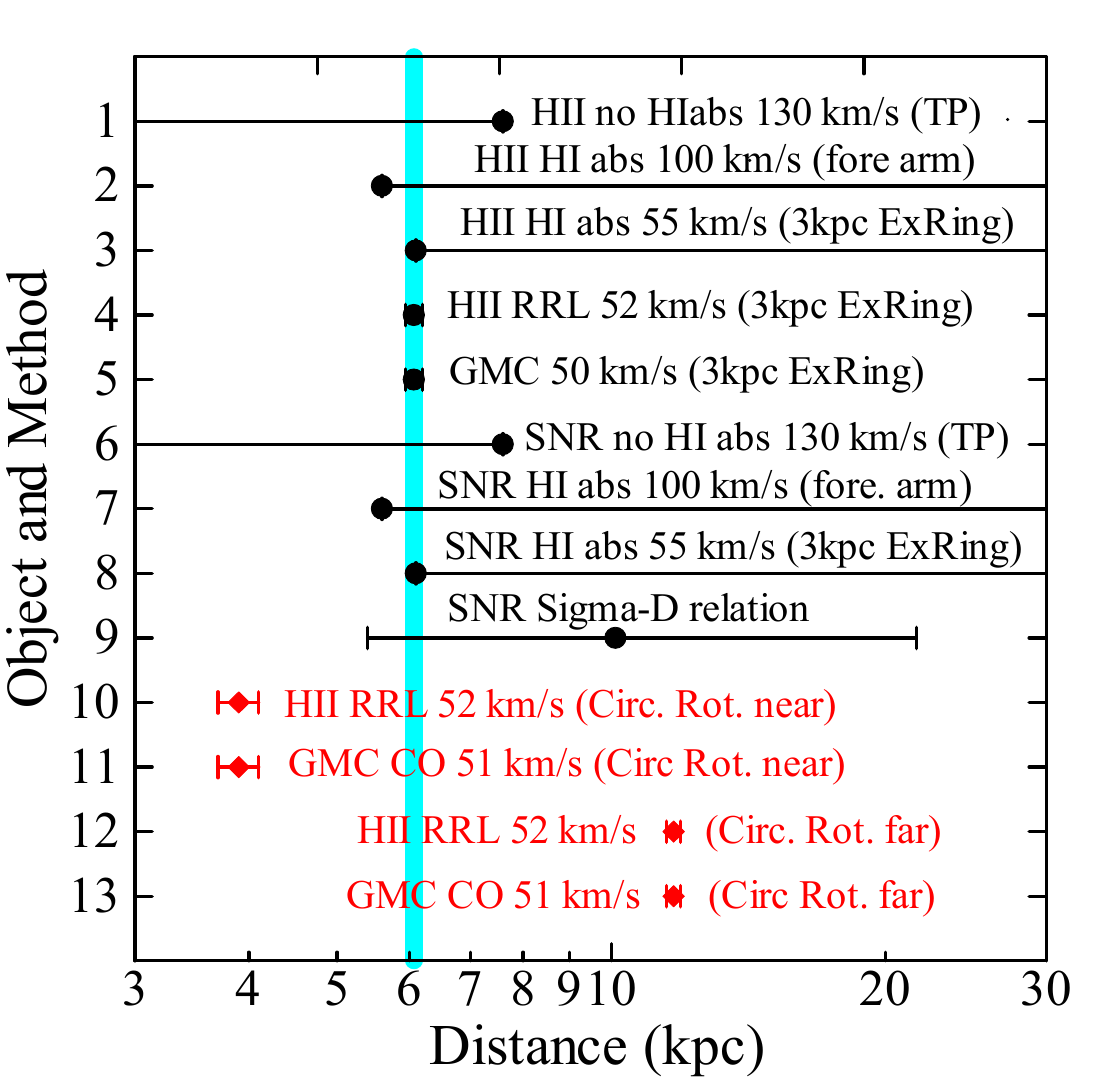}    
\end{center}
\caption{{Derived distances of the SNR, \Hii regions and GMC in G18 by various methods from Table 1. 
allowed ranges are indicated by the horizontal bars.
All the values, except for circular rotation assumption (red marks), are consistent within the allowed ranges. 
The vertical blue line indicates the distance $d=6.07\pm 0.13$ kpc to G18. 
TP stands for tangent point, and fore. arm means a foreground arm in front of the 3-kpc ER.} }
\label{fig9}  	
\end{figure}

\begin{figure} \begin{center}     
\includegraphics[width=7cm]{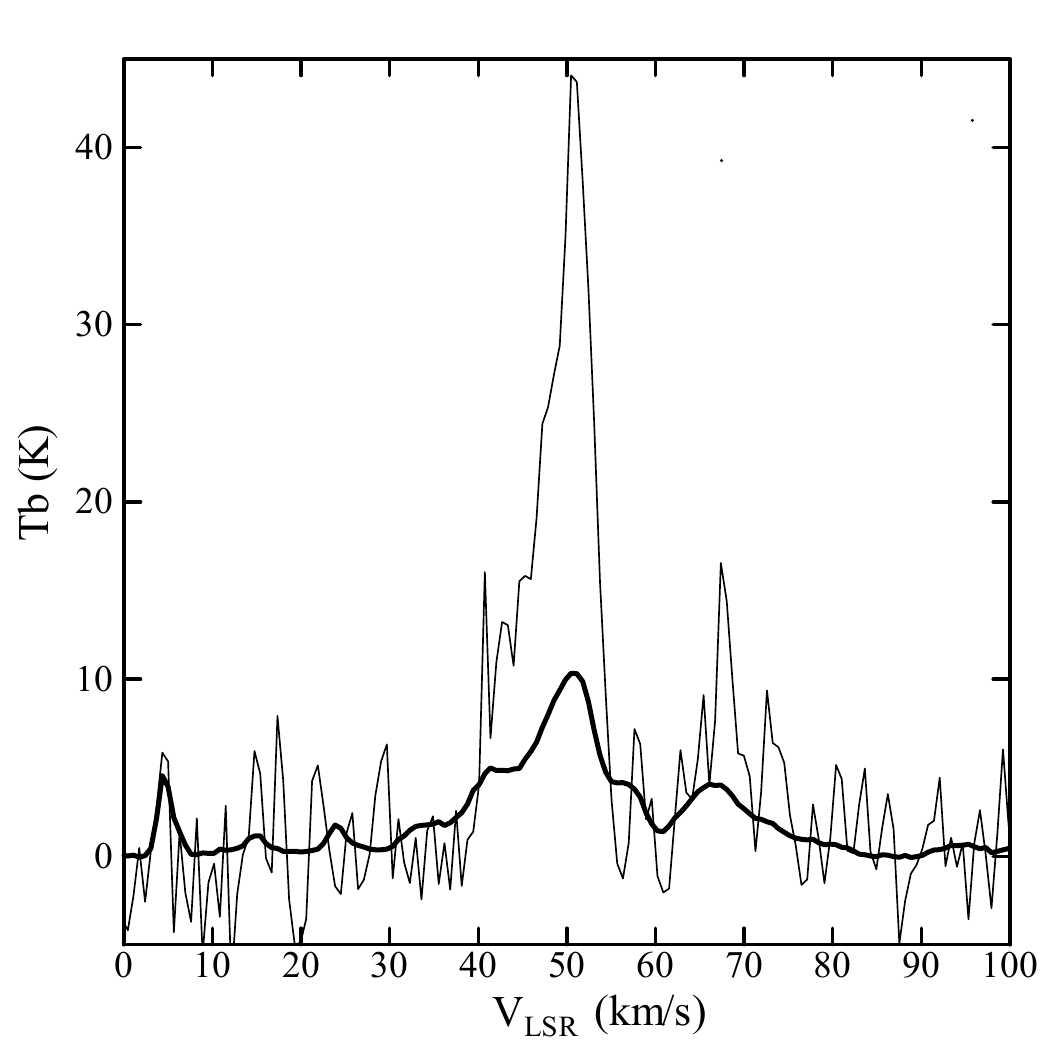}  
\end{center} 
\caption{ \co line profile at the intensity peak $(l,b)=(18\deg.10, -0\deg.32)$ by thin line  and  averaged profile within the big circle of diameter $0\deg.4$ (thick line). }
\label{fig10}  	\end{figure}

In table \ref{tab-distance} we list the possible solutions for the distance to G18, and plot them in figure \ref{fig9}.
We have obtained four possible distances: \\
(a) If the GMC and \Hii regions are rotating on a circular orbit, we have  $d_{\rm MC,\ near}=3.9\pm 0.2$ kpc as the near solution. \\
(b) Or, $d_{\rm MC, \ far}=11.7\pm 0.2$, as the far side solution. However, this is not allowed, because no \Hi absorption is observed at the terminal velocity \citep{2022ApJ...925...60D}. \\
(c)\Hi-line absorption of radio continuum emission at $\vlsr=55$ \kms by the 3-kpc ER \citep{2014MNRAS.438.1813L} indicates a distance of $d\ge 6.1$ kpc. \\
(d) If MC and \Hii regions are located on the 3-kpc ER, we uniquely obtain $d_{\rm MC}= 6.07\pm 0.13$ for the near side solution with approaching (negative) radial velocity. \\
(e) The SNR's distance is estimated by the $\Sigma-D$ relation as $d_{\rm SNR}\sim 10.1^{+11.5}_{-4.7} $ kpc, which allows a minimum distance of 6.1 kpc.  
 
We thus obtain a unique combination of GMC, \Hii and SNR distances consistent with each other, (b) + (c) + (d) + (e), but (a) is not favoured.
{Figure \ref{fig9} demonstrates the consistency of the solutions, except for those assuming the circular rotation of G18.}
The compact and isolated appearance of G18 in the LVD is also particular, showing an exceptionally high brightness clump among the others in Fig. \ref{fig8}. 
Another point is the LV location, where G18 is located near the intersection of the 3-kpc ER and 4-kpc MR, but displaced from the latter.    
From these facts, we conclude that the distance to G18 is $d=6.07$ kpc at a Galacto-centric distance of $\Rgc=3.1$ kpc.

\begin{table*}    
\begin{center}
\caption{Distances to G18.15-0.30} 
\begin{tabular}{lllll}
\hline
\hline
Object/method &Quantity & Result& Remark \\  
\hline   
HII region G18  \\ 
HI-line absorption or not & No \Hi abs. at term. velo.& $d\le 7.6$ kpc &HI abs. \citep{2009ApJS..181..255A}\\ 
 &HI abs. at 100 \kms & $d\ge 5.6\pm 0.4$ kpc &HI abs. by foreground disc\\ 
 &HI abs. at 55 \kms & $d\ge 6.1$ kpc &HI abs. by 3-kpc ER\\ 
\hline
GMC \& \Hii regions \\
Kin. dist. by circ. rot. curve&Radial velocity & $\vlsr=51\pm 2$ \kms &(1)  \\
& Galactic rot. velo. & $\Vrot=230\pm 8$ \kms&(2)\\
& GC distance &$R=4.7 \pm 0.2$ \kms \\
& Distance, near  &$d_{\rm near}=3.9\pm 0.2$ kpc &Allowed.\\
& Distance, far &{ $d_{\rm far}=11.7 \pm 0.2$ kpc }&Not allowed by no \Hi tangent abs.\\ 
\hline
GMC \& \Hii regions \\
Kin. dist. 3-kpc expa. ring &Radial velocity & $\vlsr=51\pm 2$ \kms &\\
& Galactic rot. velo. & $\Vrot=230\pm 8$ \kms&\\
& Expanding motion & $V_{\rm expa}=50$ \kms& at $l=0\deg$\\
& GC distance & $R=3.07\pm 0.14$ kpc\\
& Distance & $d_{\rm 3kpc}=6.07\pm 0.13$ kpc &Allowed.\\
\hline
SNR G18.15-0.17\\
HI-line absorption or not &No \Hi abs. at term. velo.& $d_{\rm SNR}\le 7.6$ kpc\\
 &HI abs at 100 \kms& $d_{\rm SNR}\ge 5.6\pm 0.4$ kpc & agrees with \citep{2014MNRAS.438.1813L}\\
 &HI abs at 55 \kms& $d_{\rm SNR}\ge 6.1$ kpc & \Hi abs. by 3 kpc ring\\
$\Sigma-D$ relation &&& \citep{2013ApJS..204....4P} \\
 &Angular diameter& $\theta=6.85$ arcmin & (3)\\ 
&Surface brightness& $\Sigma_{\rm 1 GHz}=3.21\times 10^{-20} \ {\small \rm w \ m^{-2} sr^{-1}}$ &(3)  \\   
&Diameter by $\Sigma-D$ & $D=20.1^{+22.9}_{-9.3}$ pc&(4)\\
 &Distance  &$d_{\rm SNR}=10.1^{+11.5}_{-4.7}$ kpc&  \\   
\hline 
\end{tabular} \\ 
\end{center}
(1) Molecular line velocity coincides with the H110$\alpha$ line velocity of $\vlsr=52$ \kms \citep{1980A&AS...40..379D,2009ApJS..181..255A}.  \\
(2) $V_0=238$ \kms, $R_0=8.2$ kpc, and the new rotation curve of the 1st quadrant of MW by FUGIN \co data \citep{2021PASJ...73L..19S}.\\
(3) Angular diameter and 20-cm brightness were measured on the VLA 20-cm map using VGPS survey \citep{2006AJ....132.1158S}, and converted to  $\Sigma_{\rm 1 GHz}$ for $\alpha=-0.6$.   
(4) $\Sigma-D$ relation for Galactic SNRs \citep{2013ApJS..204....4P}.   
\label{tab-distance}
\end{table*}  


We thus obtain a unique combination of GMC, \Hii and SNR distances consistent with each other is (b)+(c)+(d)+(e), and (a) is not favoured.
Figure \ref{fig9} demonstrates the consistency of the solutions, except for those assuming the circular rotation of G18.
The compact and isolated appearance of G18 in the LVD is also particular, showing an exceptionally high brightness clump among the others in figure \ref{fig8}. 
Another point is the LV location, where G18 is located near the intersection of the 3-kpc ER and 4-kpc MR, but displaced from the latter.    
From these facts, we conclude that the distance to G18 is $d=6.07$ kpc at a Galacto-centric distance of $\Rgc=3.1$ kpc.


\section{Kinematics, Mass and Energy}
\label{sec4}

\subsection{Molecular mass}

Given the distance of $d=6.07$ kpc, we then estimate the molecular mass of the region enclosed by a circle centered on the peak intensity position at G18.15-0.30. 
We measure the full velocity width of half maximum of the \co line emission $\sigma_v$, and the peak brightness temperature $\Tp$ using the line profiles averaged in each circle.
Then the averaged integrated intensity inside the circle is calculated by 
\be
\Ico=\sigma_v \Tp,
\ee
which yields the mean column density of $H_2$ by
\be
N_{\rm H_2}=\Xco \Ico.
\ee
The molecular mass inside radius $r$ is estimated by
\be 
M_{\rm mol}=\mu \ 2\mH \ \Xco \Ico A,
\ee
where $\mu=1.38$ is the mean molecular weight per hydrogen including metals, $A=\pi r^2$ and $r=d\ {\rm tan} \theta_{\rm dia}/2$.
We adopted the conversion factor with radial gradient given by 
\be
\log (\Xco/{\Xco}_0)=0.41(R-R_0)/r_e
\ee
with $r_e=6.7$ kpc \citep{1996PASJ...48..275A}, which yields the conversion factor at 
$R_{\rm GC}=3$ kpc as 
$\Xco=0.96\times 10^{20}$ for the local value of
${\Xco}^0=2\times 10^{20}$
and $1.4\times 10^{20}$ \xcounit for $d=3.9$ kpc.

Table \ref{tab-virial}  lists the measured parameters and calculated masses for various radii, and figure  \ref{fig11new}  plots them.
In order to compare the results with those by the current studies, we also present the values calculated for the near distance of 3.9 kpc.
Note that the velocity width $\sigma_{\rm v}$ in this table varies rather slowly with the radius compared to the steeper increase of velocity dispersion toward the center in the moment 2 map in figure \ref{fig5}.
The reason for the difference is because the values here are means within the averaging circles, while moment 2 map indicates individual local values and is biased to pick up peaky profile component due to the cut-off brightness at $\Tb\le 5$ K, leading to smaller velocity widths.

Regardless the object's distance, the calculated mass increases with the radius, so that the total mass is not uniquely determined from the observation.
We here define the cloud's size as the region inside a circle of diameter $0\deg.4$ corresponding to $r_{\rm cloud}=22$ pc, and total mass to be $M_{\rm G18}\sim 2.8\times 10^5 \Msun$ for $d=6.07$ kpc.

The mass estimated here may be compared with that obtained in a wider area from \coth intensity \citep{2013MNRAS.433.1619P} of $5\times 10^5 \Msun$ for a distance of 4 kpc and radius 24 pc, which may be converted to $1.2\times 10^6 \Msun$ and 38 pc at our distance of 6.1 kpc.
The difference will be partially due to the adopted conversion factor, which is about a half in the present calculation, and to the defined cloud size on the sky, which is also about a half here.

\subsection{Virial Mass}

\begin{figure*}
\begin{center}    
(a) Mass ~~~~~~~~~~~~~~~~~~~~~~~~~~~~~~~ (b) Mass density\\
\includegraphics[width=6cm]{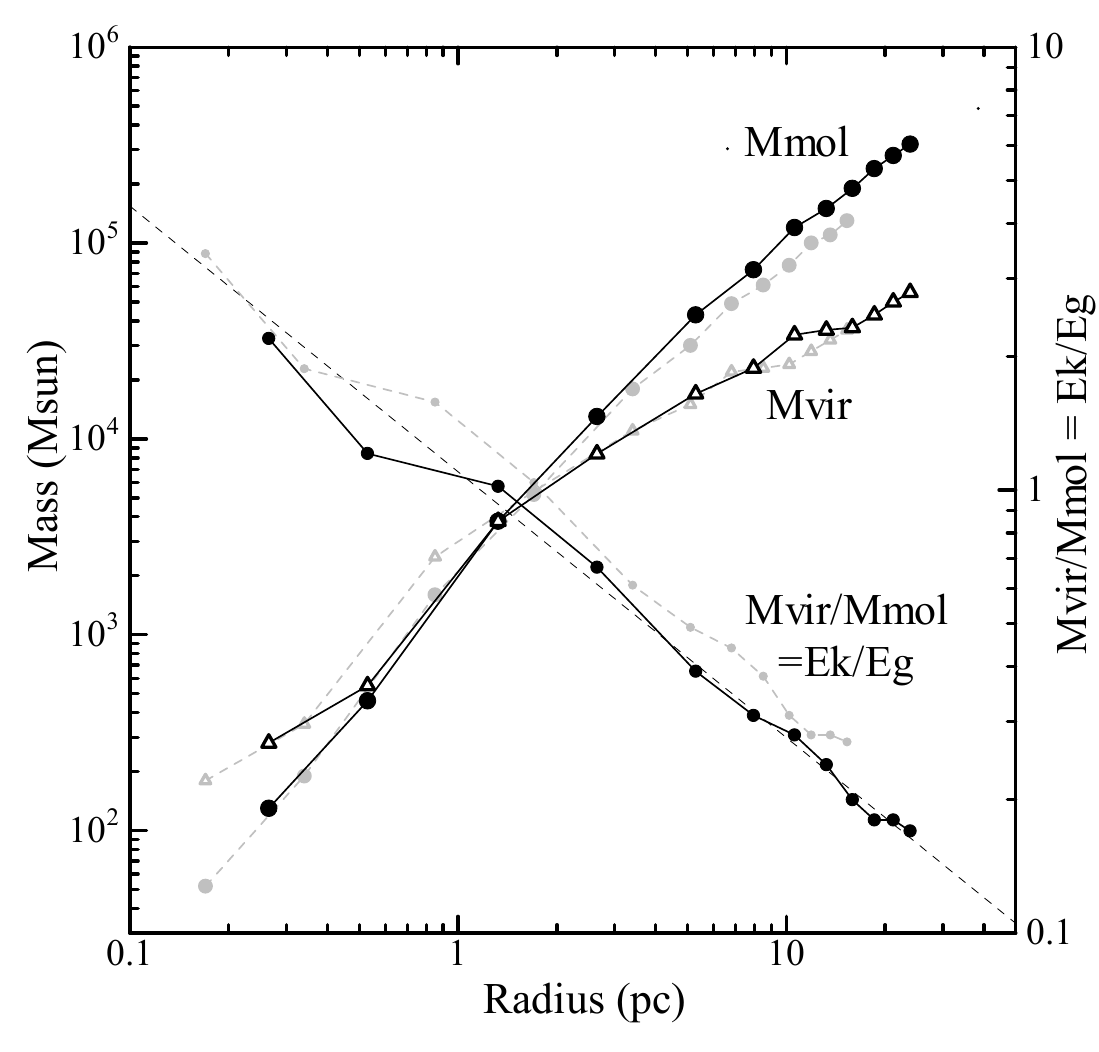} 
\includegraphics[width=6cm]{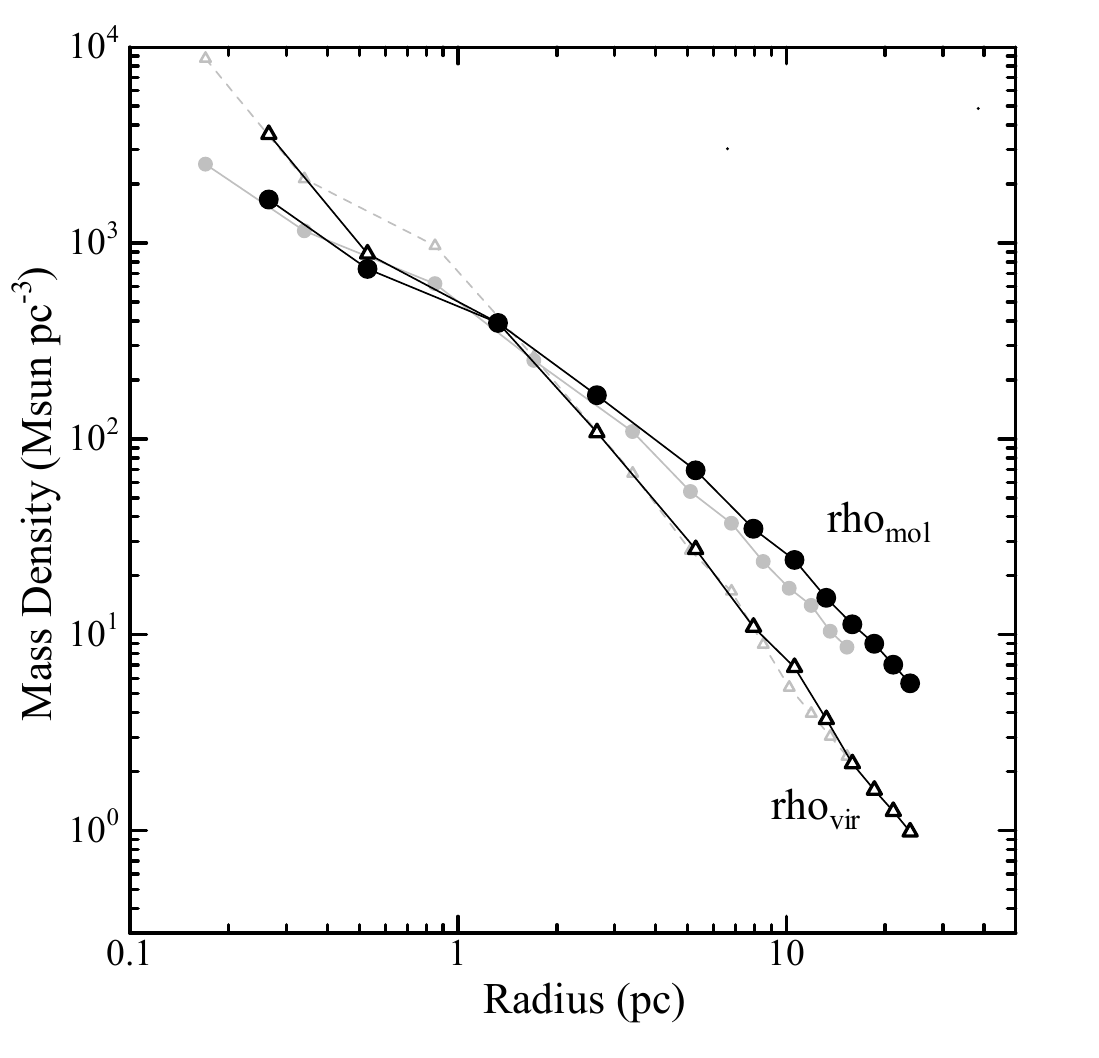}  \\   
(c) Energy ~~~~~~~~~~~~~~~~~~~~~~~~~~~~~ (d) Energy density\\
\includegraphics[width=6cm]{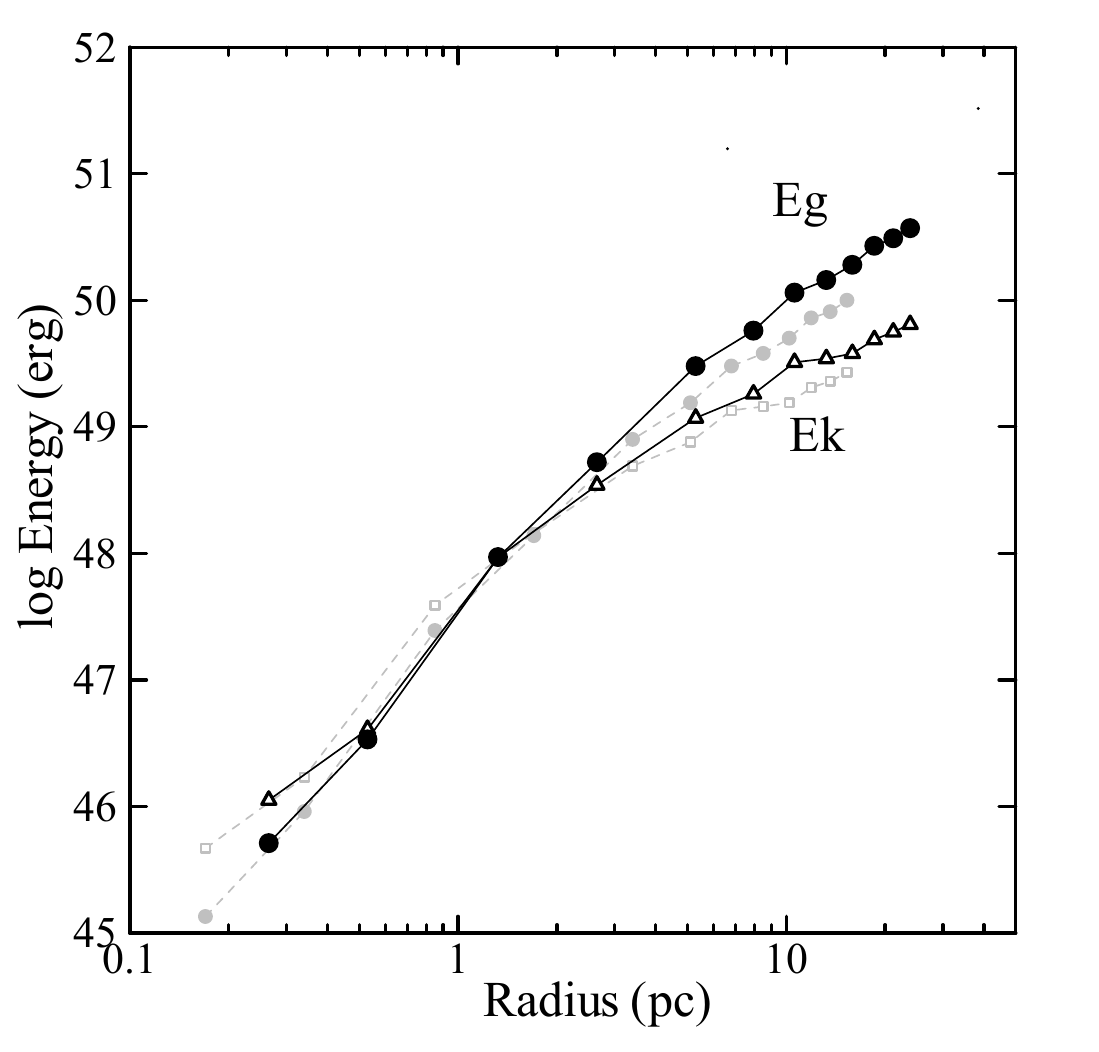}   
\includegraphics[width=6cm]{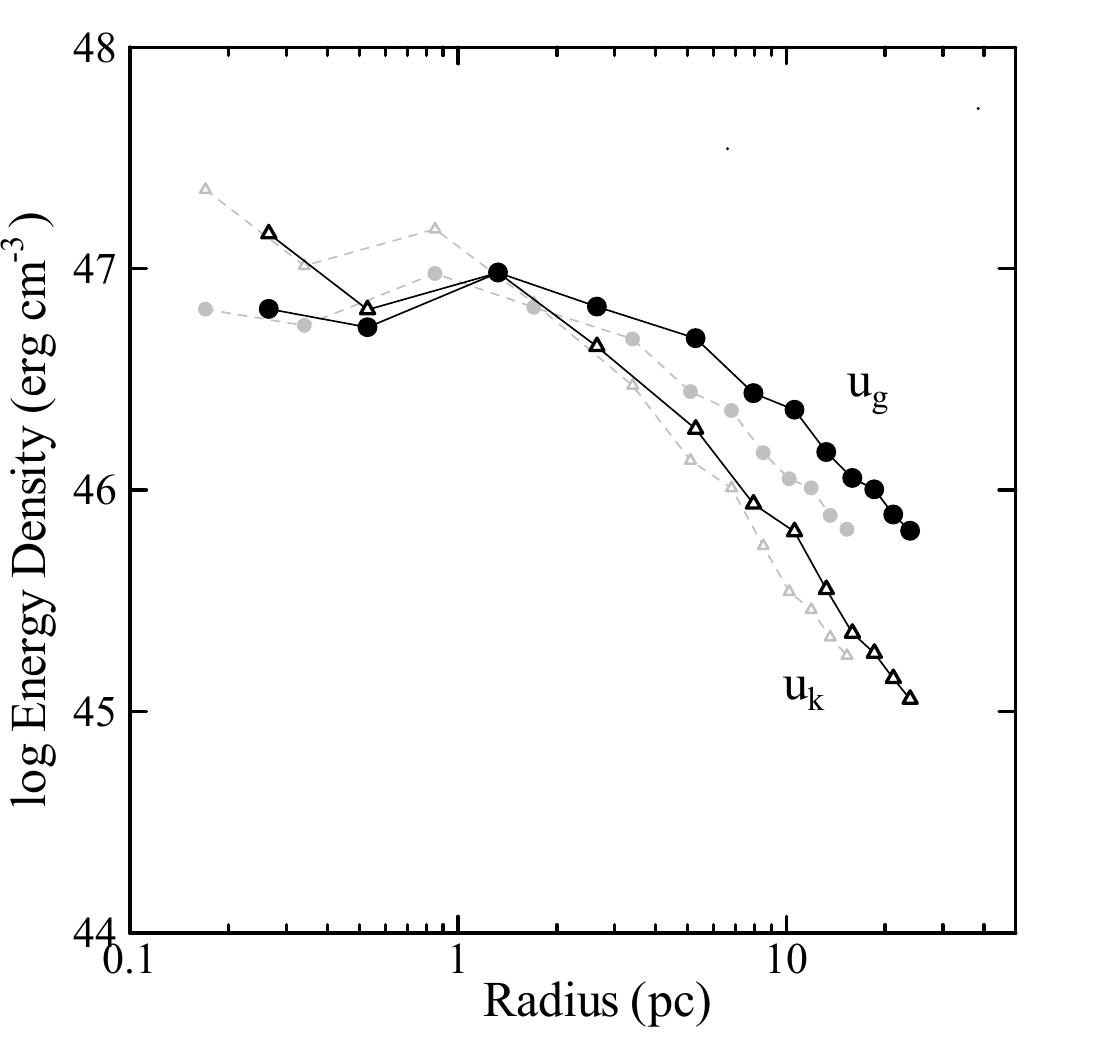}   
\end{center}
\caption{
(a) Molecular (circles) and Virial (triangles) masses within radius $r$.
Black symbols are for adopted distance of $d=6.07$ kpc.
We also plot the quantities for 3.9 kpc by the grey lines in order for comparison with the current studies.
Small symbols show the ratio of Virial to molecular masses $\mathcal{R}$, which is approximately fitted by $\mathcal{R}=1.1 (r/1 {\rm pc})^{-0.6}$.
(b) Same, but densities in $\Msun \ {\rm pc}^{-3}$.
(c) Same, but $\log \ E_k$ and $\log \ E_g$ in erg.
(b) Same, but for energy densities $u_k$ and $u_g$ in erg pc$^{-3}$.
}  
\label{fig11new}  
\end{figure*}

The half-width-of-half-maximum velocity related to the observed full velocity width $\sigma_v$ 
\be
\sigma_{\rm 3D}=1/2 \sigma_v
\ee
and radius $r$ are then used to calculated the Virial mass of the cloud inside by
 \begin{equation}
M_{\rm vir}= \sigma_{\rm 3D}^2 r/2G.
 \end{equation} 
We define the specific volume densities in a sphere of radius $r$ by
\be\rho_{\rm mol}=M_{\rm mol}/(4\pi/3 \ r^3)\ee
and
\be\rho_{\rm vir}=M_{\rm vir}/(4\pi/3 \ r^3).\ee
We plotted the calculated values against $r$ in figure \ref{fig11new} .

We introduce a ratio of the Virial to molecular masses $\mathcal{R}=M_{\rm vir}/M_{\rm mol}$.
The ratio is about $\sim 2$ in the centre, and decreases with the radius at $\mathcal{R}=4 (r/1 {\rm pc})^{-0.6}$ as indicated by the dashed line in the figure.
The cloud is more unstable near the center, and is stable, or gravitationally bound, at the edge of the cloud.
Such a radial variation may be attributed to a high rate of injection of kinetic energy in the central region.
The energy source may be the expanding \Hii shell produced by the massive star formation in the centre observed as the high-brightness \Hii region at G18.15-0.30, coinciding in position with the cloud density maximum (Fig. \ref{fig1}).
 
\subsection{Energy}

We have calculated the kinetic energy of the cloud with the derived molecular mass
\be
E_{\rm k}=1/2 M_{\rm mol} \sigma_{\rm 3D}^2
\ee
and compared it with the gravitational energy
\be
E_{\rm g}=G M_{\rm mol}^2/r,
\ee
which are also listed in table \ref{tab-virial}  along with their ratio, $E_{\rm k}/E_{\rm g}$.
We also calculated their density
\be
u_{\rm k}=E_{\rm k}/(4\pi/3 \ r^3)
\ee
and
\be
u_{\rm g}=E_{\rm g}/(4\pi/3 \ r^3),
\ee
as well as their ratio.
In figure \ref{fig11new}  we plotted their values as a function of the radius from the cloud center.
The energies of the entire cloud to the edge is on the order of $\sim 10^{50-51}$ erg,
We point out that the gravitational energy increases with the radius more steeply than the kinetic energy.
The estimated values depend on the molecular mass, and hence on the conversion factor $\Xco$, while the general trend of mutual radial variations do not change by the $\Xco$ variation within a factor of $\sim 2$. 

In the entire cloud, $E_{\rm g}$ significantly exceeds $E_{\rm k}$, or the total energy $U=E_{\rm k}-E_{\rm g}$ is negative. 
On the contrary, the core region inside $\sim 2$ pc has positive $U$, so that the gas is unstable or expanding.
From these energetics, we may consider that the GMC is gravitationally bound and contracting, whereas the central region is expanding due to some injection of kinetic energy from inside.
This property does not depend on the cloud's distance.
 
\begin{table*}   
\caption{G18.15-0.30 molecular cloud mass and energy.}
\begin{tabular}{llllllllll}
\hline 
\hline
Ang. radius & Radius $r$ & Mean $\Tb$ & $\sigma_{\rm FWHM}$& $M_{\rm vir}$& $M_
{\rm mol}$& log $E_k$& log $E_g$  & $M_{\rm vir}/M_{\rm mol}$\\ 
$(\deg)$ & (pc) &  (K) &  (\kms) & $(\Msun)$& $(\Msun)$&  (erg)& (erg) & $=E_k/E_g$ \\ 
\hline
 $d=6.07$ kpc$^\dagger$ \\
 \hline 0.005 & 0.265 & 45.0 & 6.0 & 0.28E+03 & 0.13E+03 & 46.05 & 45.71 & 2.20\\
 0.010 & 0.530 & 41.0 & 6.0 & 0.55E+03 & 0.46E+03 & 46.61 & 46.53 & 1.21\\
 0.025 & 1.324 & 32.5 & 10.0 & 0.38E+04 & 0.38E+04 & 47.97 & 47.97 & 1.02\\
 0.050 & 2.649 & 26.0 & 10.5 & 0.84E+04 & 0.13E+05 & 48.54 & 48.72 & 0.67\\
 0.100 & 5.297 & 22.0 & 10.5 & 0.17E+05 & 0.43E+05 & 49.07 & 49.48 & 0.39\\
 0.150 & 7.946 & 17.5 & 10.0 & 0.23E+05 & 0.73E+05 & 49.26 & 49.76 & 0.31\\
 0.200 & 10.594 & 15.2 & 10.5 & 0.34E+05 & 0.12E+06 & 49.51 & 50.06 & 0.28\\
 0.300 & 15.891 & 12.4 & 9.0 & 0.37E+05 & 0.19E+06 & 49.58 & 50.28 & 0.20\\
 0.400 & 21.188 & 10.3 & 9.0 & 0.50E+05 & 0.28E+06 & 49.75 & 50.49 & 0.18\\
\hline
$d=3.9$ kpc$^\ddagger$\\
\hline      
0.005 & 0.170 & 45.0 & 6.0 & 0.18E+03 & 0.52E+02 & 45.67 & 45.13 & 3.42\\
 0.010 & 0.340 & 41.0 & 6.0 & 0.35E+03 & 0.19E+03 & 46.23 & 45.96 & 1.88\\
 0.025 & 0.851 & 32.5 & 10.0 & 0.25E+04 & 0.16E+04 & 47.59 & 47.39 & 1.58\\
 0.050 & 1.702 & 26.0 & 10.5 & 0.54E+04 & 0.52E+04 & 48.16 & 48.14 & 1.04\\
 0.100 & 3.403 & 22.0 & 10.5 & 0.11E+05 & 0.18E+05 & 48.69 & 48.90 & 0.61\\
 0.150 & 5.105 & 17.5 & 10.0 & 0.15E+05 & 0.30E+05 & 48.88 & 49.19 & 0.49\\
 0.200 & 6.807 & 15.2 & 10.5 & 0.22E+05 & 0.49E+05 & 49.13 & 49.48 & 0.44\\
 0.300 & 10.210 & 12.4 & 9.0 & 0.24E+05 & 0.77E+05 & 49.19 & 49.70 & 0.31\\
 0.400 & 13.614 & 10.3 & 9.0 & 0.32E+05 & 0.11E+06 & 49.36 & 49.91 & 0.28\\
\hline 
\end{tabular}\\
$^\dagger$ G18 MC and \Hii are on the 3-kpc ER. Plotted in figure \ref{fig11new}  by black symbols for $\Xco=0.96\times 10^{20}$ \xcounit.\\
$^\ddagger$ M18 belongs to the 4-kpc MR. Plotted in figure \ref{fig11new}  by grey symbols for $\Xco=1.4\times 10^{20}$ \xcounit.
\label{tab-virial} 
\end{table*}        

The increase of kinetic energy toward the center is related to the increase of velocity dispersion.
Although the averaged values over the cloud at different radii are not so variable, the moment 2 map in figure \ref{fig5} shows an increase in the specific velocity dispersion toward the center. 

\subsection{Position-velocity diagrams}

\begin{figure} \begin{center}    
\includegraphics[height=7cm]{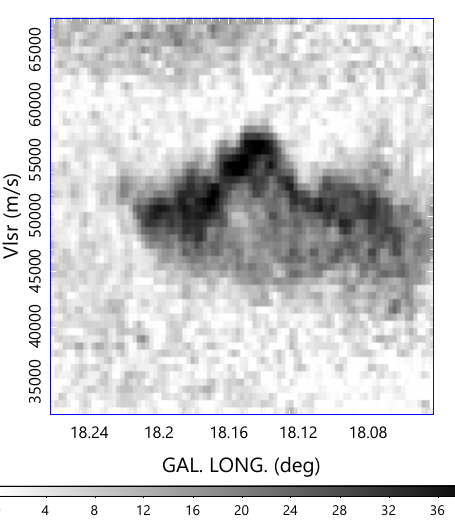}   
\includegraphics[height=6.6cm]{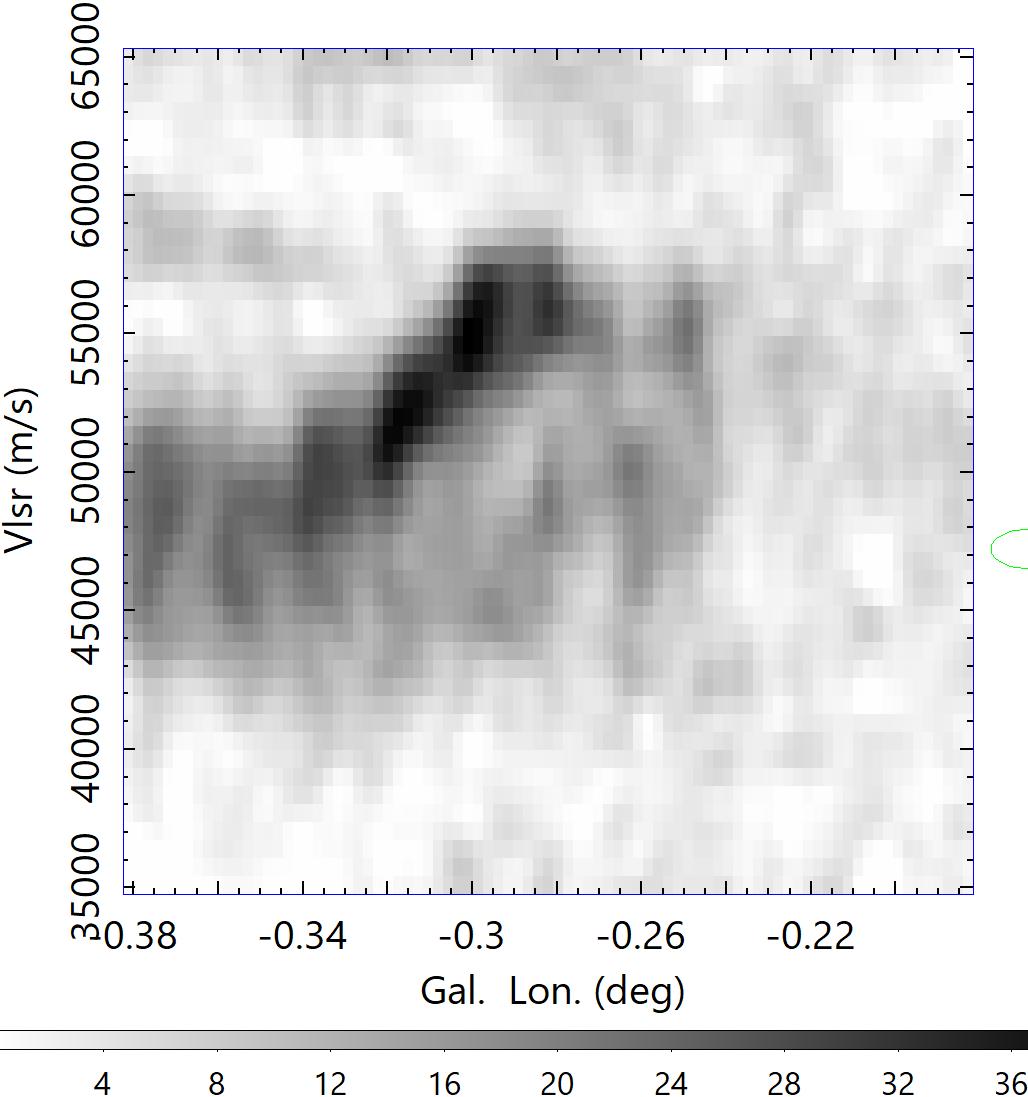} 
\includegraphics[height=6.5cm]{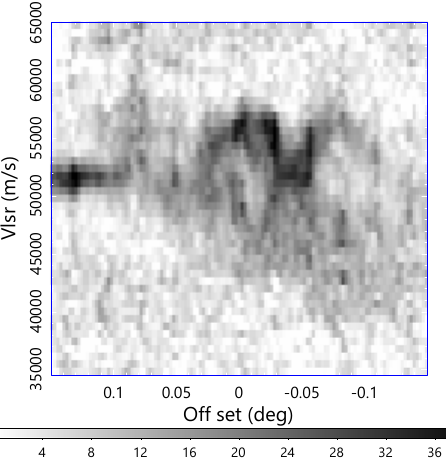} 
\end{center}
\caption{[Top] {Longitude-velocity diagram} across $b=0\deg.3$.
[Middle] {Latitude-velocity diagram} at $l=18\deg.15$. 
[Bottom] {Position-velocity diagram} at $PA=45\deg$ across G18.15-0.3, showing rotation and/or expansion of the cloud. } 
\label{fig12new}
\end{figure}

The kinematic behaviour and gaseous motion can be examined using longitude-velocity diagrams.
The right panel of figure \ref{fig2} shows \co-line LVD from FUGIN at various latitudes, and figure \ref{fig12new} enlarges the LVD across the peak intensity position of the cloud along longitude, latitude, and the major axis of the molecular hub at $PA=45\deg$.
Impressive in these figures is the oval feature symmetrically elongated in the direction of the velocity, composing an ellipse in the LVD. 
   
The elongated feature in the LVD may be explained by various mechanisms such as expanding shells due to the pressure of \Hii regions
 \citep{2013MNRAS.433.1619P}, 
cloud collision
 \citep{2022ApJ...925...60D}.
interaction of two filaments
 \citep{2020A&A...642A..87K},  
or by the gravitational contraction.  

We here notice that the LVD oval in Fig. \ref{fig12new} is extending for $\sim \pm 5 \ekms$ symmetric with respect to the systemic velocity $\vlsr=51$ \kms.
Such symmetry of velocity may be naturally attributed to an expansion of a shell rather than to a cloud collision, which is characterized by a lopsided bridge feature extending from the targeted GMC to the colliding body \citep{2015MNRAS.454.1634H}.

So, we next examine if the observed position-velocity diagrams and morphology of the {molecular bubbles} can be explained by expanding shells. 
The top panels show expanding-shell model by star-forming activity at the edge of a molecular cloud represented by the white ellipse. 
The expansion of a gaseous shell was traced by the adiabatic shock wave approximation in order to examine the qualitative behavior of the expanding {bubble} \citep{1971Ap&SS..14..431S,2022MNRAS.509.5809S}.
The left panel shows the shock fronts at two epochs at $t\sim 0.5$ and 1 My as projected on the $(x,z)$ plane, and the right shows corresponding position-velocity diagram seen from the $z$ direction of the shell.
Here, the kinetic energy is injected continuously at the coordinate origin by an injection rate of $dE/dt \sim 2\times 10^{49}$ erg My$^{-1}$.  
The parent cloud is assumed to have an off-center Gaussian density profile as indicated by the white ellipse with the center density of $10^4$ \Htwo \percc.
The parameter combination corresponds to a linear scale of the figure frame of $\sim 20$ pc and velocity frame $\sim 10$ \kms.

\begin{figure*} 
\begin{center}    
{\large Shock wave simulation: Projected on the sky (left), and position-velocity diagram (right)}\\
\includegraphics[width=6cm]{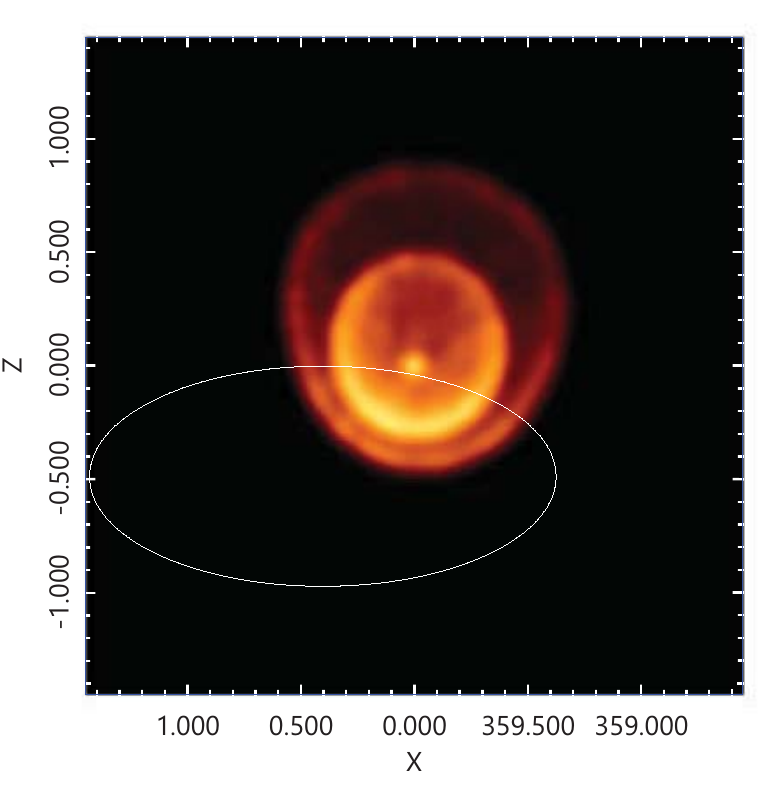}     
\includegraphics[width=6cm]{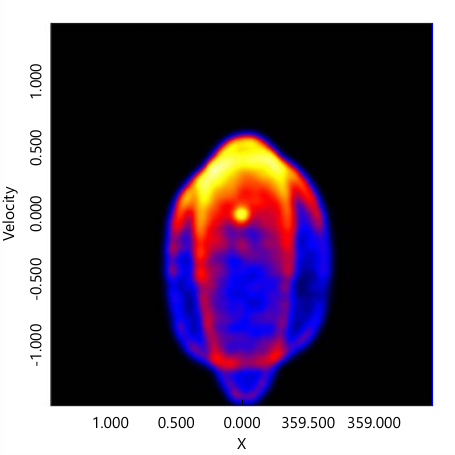}  \\
{\large $\Ico$: On the sky (37 pc $\times$ 37 pc) and LVD (37 pc $\times \vlsr$)}\\
\includegraphics[width=6cm]{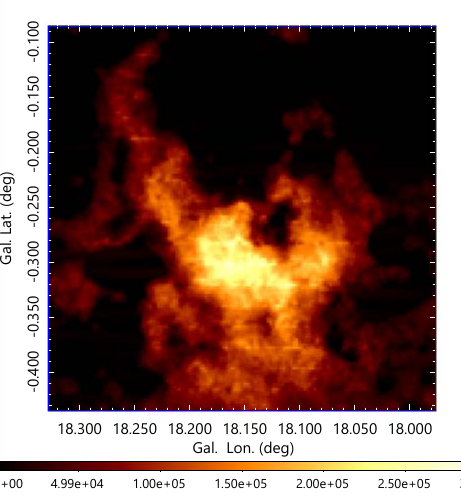}
\includegraphics[width=6cm]{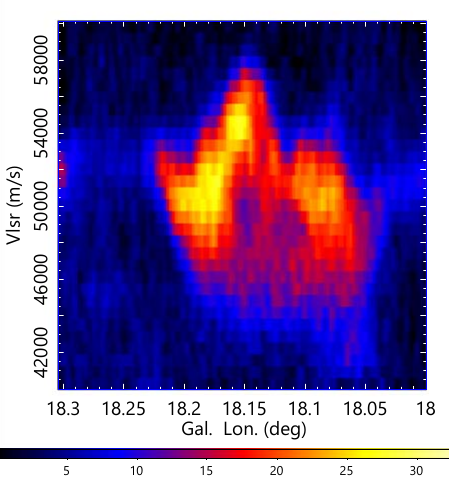}   \\  
{\large $\Tb$ at 50 \kms channel map: On the sky (19 pc $\times$ 19 pc), and LVD (19 pc $\times \pm 17$ \kms}\\
\includegraphics[width=12cm]{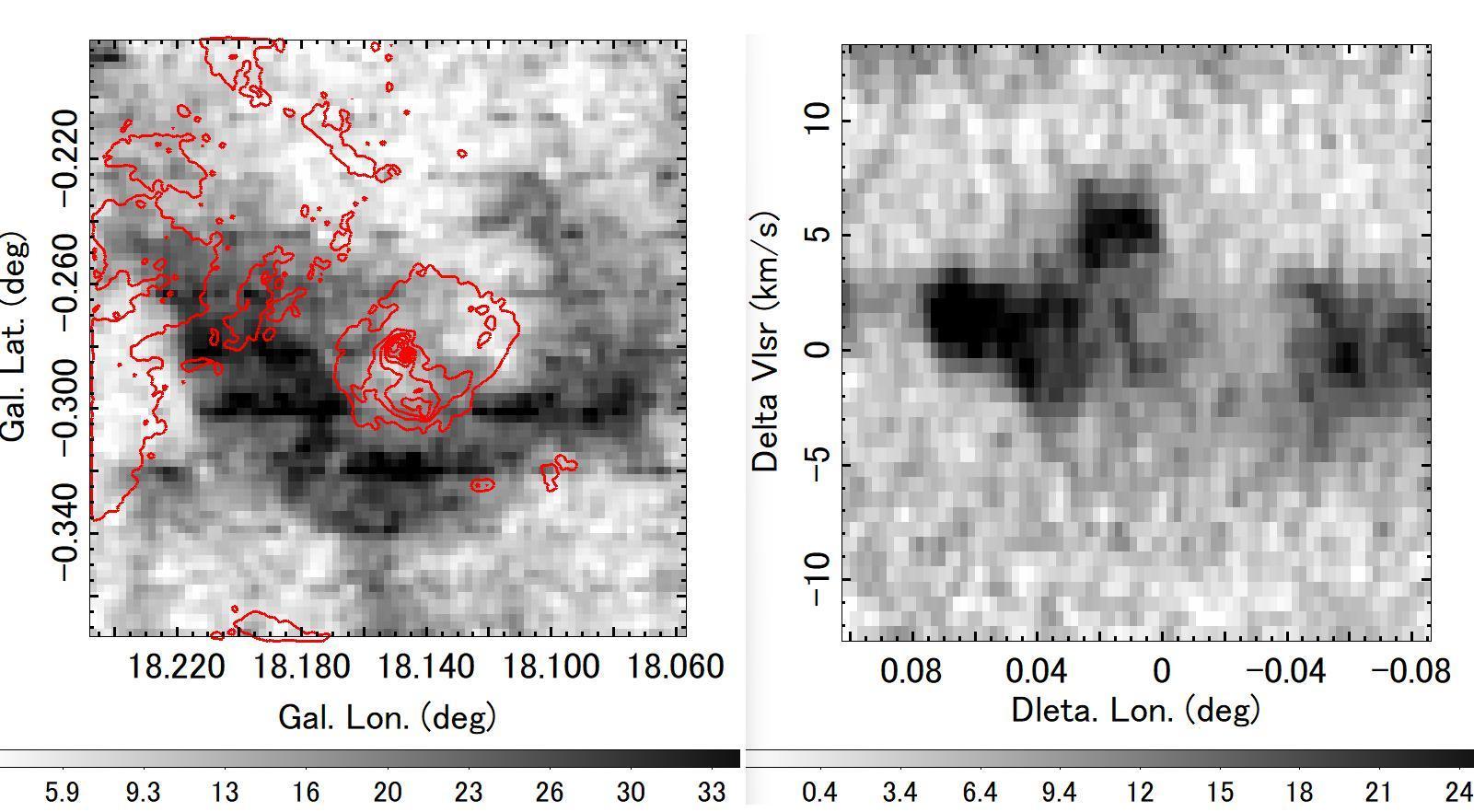}   \\
\end{center}
\caption{[Top] Left: Star-forming bubble model, showing an expanding shock front at two epochs. White ellipse illustrates the background cloud. 
Right: Corresponding position-velocity diagram seen from the bubble top. Oval as well as bridge features are reproduced. (Scale is arbitrary.)
[Middle] Left: Integrated \co intensity of G18 in K m s$^{-1}$.
Right: LVD of $\Tb$ (K) averaged between $b=-0\deg.31$ and $-0\deg.275$.
[Bottom] Left: Channel $\Tb$ map around bubble 1 at $\vlsr=50$ \kms, showing the cavity around the strong continuum source of the \Hii region shown by the contours.
Scale is enlarged by three times than the middle panels.
Right: LVD across the central \Hii region of the left panel, showing a spherical LV feature indicative of a symmetric expansion of a bubble at $\sim \pm 8$ \kms.}
\label{fig13new} 	
\end{figure*}

The middle panels show the integrated \co intensity (K m s$^{-1}$) of G18 molecular bubble1 1 to 3 and the averaged LV diagram near bubble 1 from Fig. \ref{fig12new}.
The bottom panels show an enlarged (three times than the middle panel) channel map $\Tb$ (K) at $\vlsr=50$ \kms around bubble 1 and LVD across the strong continuum source of the \Hii region shown by the contours in the left panel. The LVD shows an expanding oval feature indicative of a spherical expansion of a shell (bubble) at $\sim \pm 8$ \kms.
Such a symmetric and high velocity expansion cannot be explained by a cloud collision, which postulates a lopsided oval with a bridge feature \cite{2015MNRAS.454.1634H}.

The cavity structure opening from the cloud's edge toward the north and the {position-velocity} oval elongated in the velocity direction are well simulated by this simple model.
The bridge feature, often raised as the evidence for the collision model \citep{2021PASJ...73S...1F}, is also produced in the expanding {bubble} model. 
This model is also consistent with the location of the brightest \Hii region with a hot spot G18.144-0.282 as the energy injection source inside the cavity as observed in figure \ref{fig6}, where a couple of O stars are reported to be located \citep{2013MNRAS.433.1619P}. 
It is also pointed out that {molecular bubbles} 2 and 3 are approximately centred by \Hii region G18.19-0.18 overlapping with the SNR.
These multi-bubble structure is recognized as corresponding multi-LVD ovals in figure \ref{fig12new}.
We also stress that the outer shell-shaped \Hii regions G18.18-0.40 (N21) and G18.26-0.31 (N22) are located surrounded by CO shell or arcs.

\section{{Discussion}}
\label{sec5}
 
{We here try to explain the origin of the G18 system according to a classical scenario of star formation in a GMC based on the facts presented in this paper.
We consider a core-haloed GMC associated with expanding {molecular bubbles} driven by an OB star cluster born by the sequential star formation in the hub centre and their feedback. }

\def\htwocc{H$_2$ cm$^{-3}$}
\def\nhtwo{n_{\rm H_2}}

Due to the gravitational contraction of the spokes toward the gravity center, the density in the hub increases in the free-fall time of $t_{\rm ff}\sim 2$ My for $\nhtwo\sim 10^3$ \htwocc in the spokes.
Star formation will take place in the compressed hub center by the gravitational (Jeans) instability \citep{2013MNRAS.433.1619P}, with the Jeans mass of $M_{\rm J}\sim 9 \Msun$ and growth time of $t_{\rm J}\sim 0.7$ My for the observed density of $n_{\rm H_2}\sim 10^4$  \htwcc(Fig. \ref{fig11new}) and assumed temperature of $T\sim 20$ K.
Namely, massive (OB) stars greater than $9 \Msun$ are born in a relatively short time scale of $t_{\rm J} \sim 1$ My. 
Although unresolved in the present observations, the hub center would be turbulent, full of higher density clumps, which would in parallel leads to the formation of lower Jeans mass stars in shorter time scales.
Thus, once the hub is compressed to $\sim 10^4$ \htwocc, a star cluster including massive to low mass stars are born within a time scale of $\sim 1$ My.

Energy injection by the OB stars produces an \Hii region and compresses the surrounding molecular gas to induce the next-generation stars following the sequential star formation scenario \citep{1977ApJ...214..725E,2013MNRAS.433.1619P}, as indeed observed in the bottom panel of Fig. \ref{fig6}.
The expanding velocity of the innermost bubble 1 surrounding the \Hii region is about $v_{\rm exp}\sim \pm 7$ \kms according to the LVD in Fig. \ref{fig8}.
Corresponding compression time of the high-density molecular hub is, therefore, $t_{\rm comp}\sim r_{\rm bub}/v_{\rm exp}\sim 0.4$ My for $r_{\rm bub}\sim 2.6$ pc.

At the same time the extruded molecular gas expands and blows out to the lower density side of the hub between the spokes, and creates the observed {bubbles} 1 to 4.
Fig. \ref{fig13new} shows a simple simulation of such expanding shell producing an elongated LVD.
The expanding bubbles, then compress the neighboring spokes \citep{2015A&A...580A..49I}, making the spokes denser and tighter.  

A part of the blown off molecular gas further expands outward, but it is stopped by the surrounding ISM and fed back to the spoke and hub by the gravity.
The fed-back gas is gathered through funnels along the spokes and returns to the hub
 \citep{2013A&A...555A.112P,2018A&A...613A..11W,2019A&A...629A..81T}. 
Such a funnel flow will cause velocity gradient as observed in the velocity field in Fig. \ref{fig4}.
The outermost regions of the spokes with large cross sections increase the chance of capture of surrounding ISM, supplying more gas to the system.

\begin{figure} 
\begin{center}

\includegraphics[width=8cm]{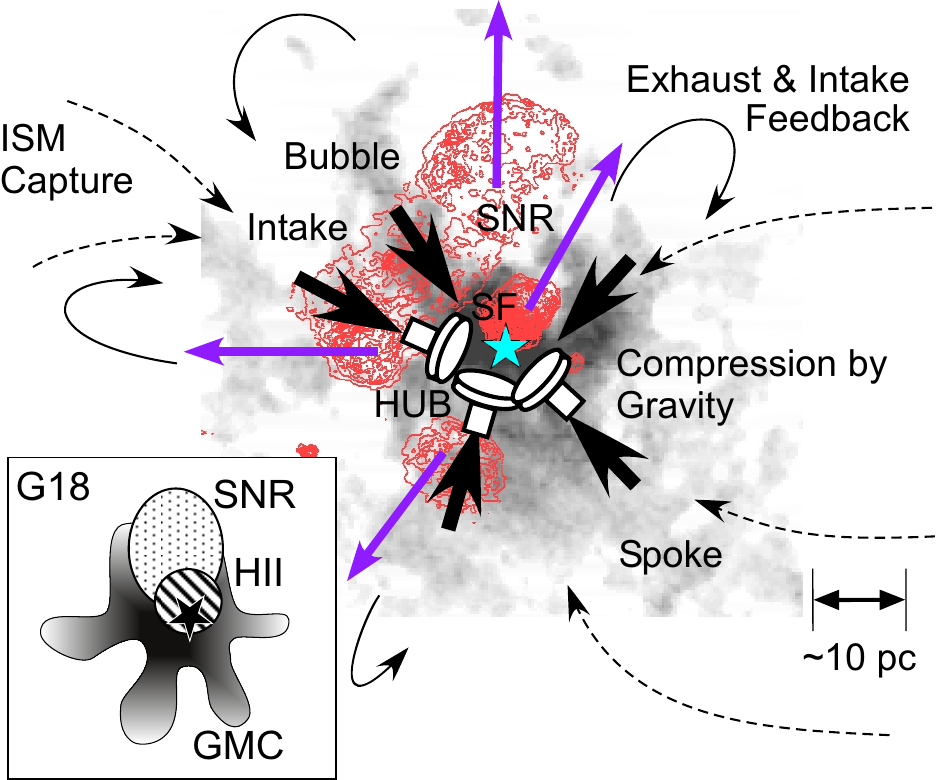}  
\end{center}
\caption{ICE (internal combustion engine) model for sustainable star formation in a GMC with molecular hub-spoke (filament)-bubble structure.}
\label{fig14new} 	
\end{figure}

The hub, spokes (filaments) and bubbles therefore constitute an internal combustion engine (ICE) as illustrated in Fig. \ref{fig14new}.
Gas in-taken from the ISM through the spoke manifold is compressed into the hub by the gravitational force as the piston, explodes by the nuclear energy released by the newly born massive stars, expands and is exhausted through the low-density region between the spokes. 
G18 thus constitutes a standalone sustainable system around the GMC with hub-spoke (filament)-bubble structure with continuous SF activity.
The SF is continuous but can be sporadic on a time scale of one cycle of about the free-fall time in the spokes of $\sim 1$ My.
This time scale is on the same order as that required for the triggered formation of massive stars by the cloud-collision scenario \citep{2021PASJ...73S...1F}, by gravitational construction of the hub from multiple filaments \citep{2009ApJ...700.1609M}, or the sequential star formation \citep{1977ApJ...214..725E}.

\section{Conclusions} 
\label{sec6}

We have obtained detailed analysis of the morphology and kinematics of the GMC G18.15-0.30 associated with multiple \Hii regions and an SNR.
The cloud comprises a dense central cloud making a hub, several bubbles, and X-shaped spokes extending radially for more than $\sim 0\deg.6$ ($\sim 50$ pc). 
 
G18 is located on the LVD at $\vlsr=51$ \kms close to the intersection of the 4-kpc molecular ring and 3-kpc expanding ring.  
Carefully determining the distance ranges for the GMC, \Hii regions and SNR using the velocities of CO and \Hii recombination lines,\Hi absorption line, and $\Sigma-D$ relation for SNR, we concluded that the distance to G18 is $6.1$ kpc in the 3-kpc expanding ring.
We argue that the SNR G18.15-0.17 is physically associated with G18 system, composing a triple association of GMC, \Hii region, and SNR.

Based on the derived distance We determined the physical parameters such as the mass and energy of the GMC, and considered their behaviours inside the cloud (table \ref{tab-virial}  and figure \ref{fig11new}). 
We comment that the result is dependent on the CO-to-H$_2$ conversion factor, which include uncertainty of a factor of two.
However, the relative variation between the quantities led to the following physical conditions in the cloud.

In the cloud center within a few pc, the kinetic energy exceeds over the gravitational energy, indicating that the core is dynamic and expanding.
In the outer region from the outer hub and spokes, it is vise versa, indicating that the system is stable and gravitationally bound to the system.  

The dynamic state of the central region is explained by the energy injection expanding motion due to star forming activity. 
The entire cloud system is gravitationally stable, and the star forming activity in the hub is sustained by the feedback of the exhausted gas as well as by capture of interstellar gas through the longitudinal flow along the spokes (filaments), and composes an internal combustion engine  (Fig. \ref{fig14new}).

Finally, we comment that, if the location in the 3-kpc expanding ring is the case, G18 would be the first example of a star forming activity in a strongly shocked arm induced by the Galactic Centre explosion or by a deep bar potential.
In this context we stress the absence of the molecular disc in the upper half of figure \ref{fig3} at $\vlsr \sim  50$ \kms, which might be related to the sweeping of the upper disc by the expanding shock wave from the GC producing the North Polar Spur \citep{2021MNRAS.506.2170S}.

\section*{Acknowledgments}
The author is indebted to the FUGIN team for the CO data observed with the Nobeyama 45-m telescope operated by the NAOJ (National Astronomical Observatory of Japan).

\section*{Data availability}
The FUGIN data were retrieved from the JVO portal at {http://jvo.nao.ac.jp/portal}, and
MAGPIS data from https://third.ucllnl.org/cgi-bin/colorcutout. 

\section*{Conflict of interest}
The author declares that there is no conflict of interest.


\begin{appendix}
\section{{A distant molecular cloud with compact \Hii region G18.305-0.391+32}} \label{appendix}

\begin{figure*} 
\begin{center}
$\vlsr=29.7$ \kms \hskip 1.5cm $32.3$ \kms ~~~~~~~\\
\includegraphics[width=5cm]{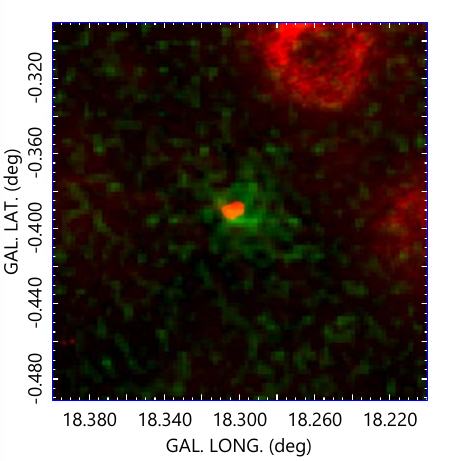}  
\includegraphics[width=5cm]{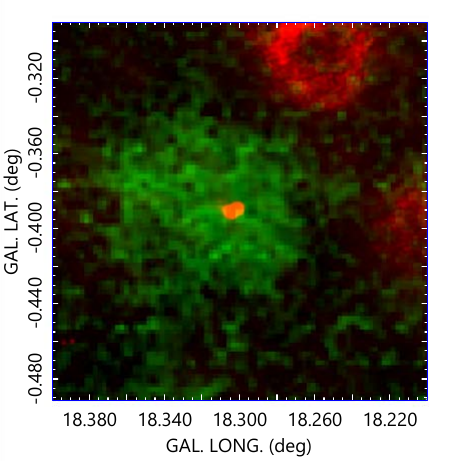} \\
$34.2$ \kms \hskip 1.5cm $34.8$ \kms \\
\includegraphics[width=5cm]{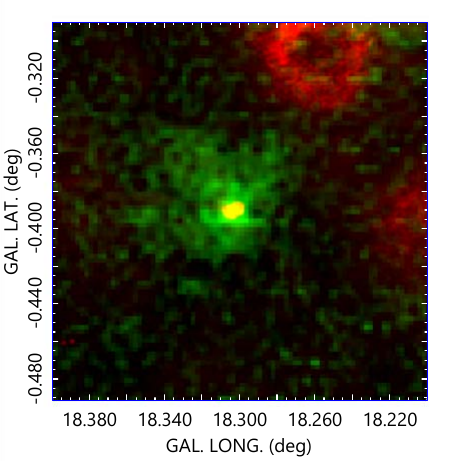} 
\includegraphics[width=5cm]{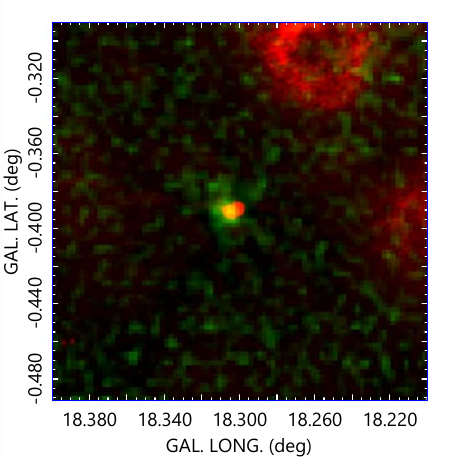} 
\end{center}
\caption{{G18.305-0.391+34 in 20-cm continuum (red: 0 to 0.2 Jy beam$^{-1}$) and \co\ brightness (green: 0 to 30 K). The source is not related to G18, and the kinematic distance  is 12.6 kpc.}}
\label{fig-app} 	
\end{figure*}

{The compact \Hii region at G18.305-0.391 with the RRL velocity of 32 \kms \citep{1980A&AS...40..379D} is located in the center of a core-haloed molecular cloud at radial velocity $\vlsr=34$ \kms in the FUGIN CO data cube.
Fig. \ref{fig-app} shows close up maps of this source in 20-cm continuum and \co\ brightness at some different velocities.
The kinematic distance to the molecular cloud is determined to be $5.4\pm 0.4$ and $d=12.6\pm 0.5$ kpc for near- and far-side solutions, respectively.  
Therefore, the source is not related to the G18 system discussed in this paper.
The peak $\Tb$ in the \co\ line is 30 K at the core, and the averaged temperature within $0\deg.1$ diameter circle is 11 K.  
If we adopt the far distance, the molecular mass is estimated to be $\sim 7\times 10^4 \Msun$.
}

\end{appendix}
\end{document}